**Thrust Based On Changes In Angular Momentum**

Yury N. Razoumny*, Sergei A. Kupreev
Peoples Friendship University of Russia (RUDN University)
Email: razoumny-yun@rudn.ru, kupreev-sa@rudn.ru
Address: 6 Miklukho-Maklaya Street, Moscow, 117198, Russian Federation

**Highlights**

1. Demonstration of orbital motion control through kinetic moment variation
2. Theoretical basis for quantum-scale momentum control using particle spin
3. Spin-based propulsion efficiency exceeds conventional methods by orders of magnitude
4. Thrust hypothesis via asymmetric low-energy particle exchange
5. Experimental vacuum tests confirm anomalous effects predicted by theory

**Abstract:**

A novel non-reactive thrust principle based on controlling the angular momentum of a material body is proposed. Theoretically, it is shown that asymmetric emission/absorption of low-energy particle fluxes with spin in a direction perpendicular to the motion enables the creation of a propulsion system whose energy efficiency exceeds that of a photon engine by several orders of magnitude near massive bodies. Using the example of a "dumbbell-flywheel" system dynamics in a central gravitational field, the possibility of controlling orbital parameters without propellant consumption is demonstrated. Experiments in a vacuum chamber revealed anomalies consistent with the hypothesis of spacetime spin polarization. The developed approach offers a mechanistic interpretation of wave-particle duality and suggests new pathways for unifying gravity with quantum mechanics. The obtained results open prospects for the development of next-generation propulsion systems.

## 1. Introduction

The further expansion of humanity into space is constrained by a fundamental limitation inherent in the reactive motion principle formulated by K.E. Tsiolkovsky [1]: the mandatory expenditure of propellant. This key limitation leads to extreme resource intensity for large-scale interplanetary missions and makes flights beyond the Solar System practically unfeasible at current technological levels. Existing developments in advanced propulsion systems, such as solar sails, ion or plasma thrusters, only partially mitigate this problem, as they are either characterized by low thrust magnitude or ultimately remain within the reactive paradigm.

As an alternative, research is being conducted aimed at utilizing gravitational field energy for movement and maneuvering [2, 3]. Although some proposed projects in this area have faced criticism from the scientific community, the fundamental idea of searching for radically new, more efficient methods of thrust generation remains relevant [4]. The approach to solving the thrust problem proposed in this work is based on new physical principles and has been developed in strict accordance with the fundamental conservation laws: momentum, angular momentum, energy, and the position of the system's center of mass.

The ideas of the controlled motion of a body in the central gravitational field without mass consumption were put forward by specialists in the field of dynamics of orbital tether systems [5-10]. In [5, 6], V.V. Beletskiy proposed the method and model of a spacecraft in the form of a dumbbell, capable of making space flights between coplanar orbits without consuming a working fluid. A large-sized dumbbell is located in space along the binormal to the orbit so that its center of mass moves along the orbit, in the plane of which the attracting center is located, and the end masses are on opposite sides of this plane. It is shown that by changing the length of the dumbbell bar it is possible to increase the eccentricity of the orbit.

In [7, 8], the dynamical behavior of a tethered connected satellite system during tether length variation is considered. It is shown that appropriate length variation laws can be used to modify the characteristics of the assumed elliptical orbit of the system mass center, as well as to solve the problem of delivering cargo from orbit without consuming fuel.

The monograph [9] proposes control schemes for orbital elements due to different orientations of a dumbbell with a variable bar length, including the use of flywheels to hold the dumbbell in a given position. The idea of using a rotating orbital tether system with a variable bond length is proposed, which is the fact that, due to internal forces, the distance between the end bodies changes and thereby the angular velocity of rotation of the system is controlled so that the system is in the desired orientation longer than in the position, giving the opposite effect of control.

In [10], the orbital elements are controlled by a tether system with a periodically varying length by taking into account the inhomogeneity of the gravitational field.

The internal logic of the development of science prompts to take into account fundamental research in the field of quantum mechanics. The study of the motion essence of material bodies on the basis of the fundamental laws of classical and quantum mechanics opens horizons for a broader understanding of the phenomena of physics and, in particular, for the formation of ideas for creating thrust based on new physical principles.

In this context, a relevant task is the search for and justification of fundamentally new physical principles for thrust generation that are not associated with mass ejection. One such promising direction is the control of the intrinsic angular momentum (kinetic moment) of a physical system. Classical mechanics recognizes effects demonstrating the relationship between rotational and translational motion in non-uniform force fields, such as in the dynamics of tether systems. However, these macroscopic effects have traditionally been considered extremely weak for practical application in propulsion systems.

This work theoretically and experimentally investigates a pathway to overcome this limitation by transferring the principle of angular momentum control to the quantum level. It is shown that asymmetric emission or absorption of elementary particles with spin (intrinsic angular momentum) in a direction perpendicular to the body's motion can lead to significant thrust generation. Unlike rocket engines, this approach utilizes not the particle's momentum but its spin, which, as proven in this work, provides orders of magnitude higher energy efficiency near massive bodies.

The aim of this article is to provide a systematic substantiation of a new method for thrust generation based on kinetic moment control. To achieve this goal, the following tasks have been sequentially accomplished:

1. Using the classical "dumbbell-flywheel" system as an example, the fundamental possibility of controlling orbital motion through changes in intrinsic angular momentum has been demonstrated.

2. Extrapolation of this principle to the quantum level has been conducted, along with a comparative assessment showing the potential energy advantage of using elementary particle spin compared to reactive methods.

3. A hypothesis linking accelerated body motion with asymmetry in flows of low-energy particles with spin has been formulated, and a mechanistic model of space-time spin polarization has been proposed.

4. Results from pilot experiments in a vacuum chamber have been presented, revealing anomalies in test body motion consistent with the hypothesis predictions.

5. A strategy for further theoretical, experimental, and applied research has been proposed.

The scientific novelty of the work lies in its interdisciplinary approach, connecting classical orbital mechanics with quantum principles to substantiate a new method of thrust generation. The practical significance consists in forming a theoretical basis and roadmap for developing next-generation propulsion systems whose energy efficiency could exceed existing analogues by orders of magnitude. The development of this direction also contributes to deepening understanding of fundamental physics problems such as the nature of inertia, wave-particle duality, and gravitational interaction. The publication of this work is intended to stimulate broad interdisciplinary discussion and attract the scientific community's attention to the prospects of developing reactionless thrust based on angular momentum control.

## 2. Classical Mechanical Analogy: Principle Demonstration

This section serves as the starting point for developing a research strategy. Using a well-studied example from classical dynamics, it demonstrates a key principle: changing a system's internal angular momentum can alter its translational motion in a non-uniform force field. This principle, observed at the macroscopic level, is proposed as an analog for transfer to the quantum level.

### 2.1. Dumbbell Motion Model in a Central Gravitational Field

In the central gravity field, there is a relationship between rotational motion relative to the center of mass of the body and the radial motion of the body [11, 12].

Consider the movement of a rigid dumbbell in the Earth central gravitational field. Suppose that two finite exact masses of a dumbbell $m_1$ and $m_2$ are connected by a weightless rigid rod $D$ ($D \ll r$). Two external forces of attraction $\boldsymbol{G_1}$ and $\boldsymbol{G_2}$ (Fig. 1) are acted upon the dumbbell.

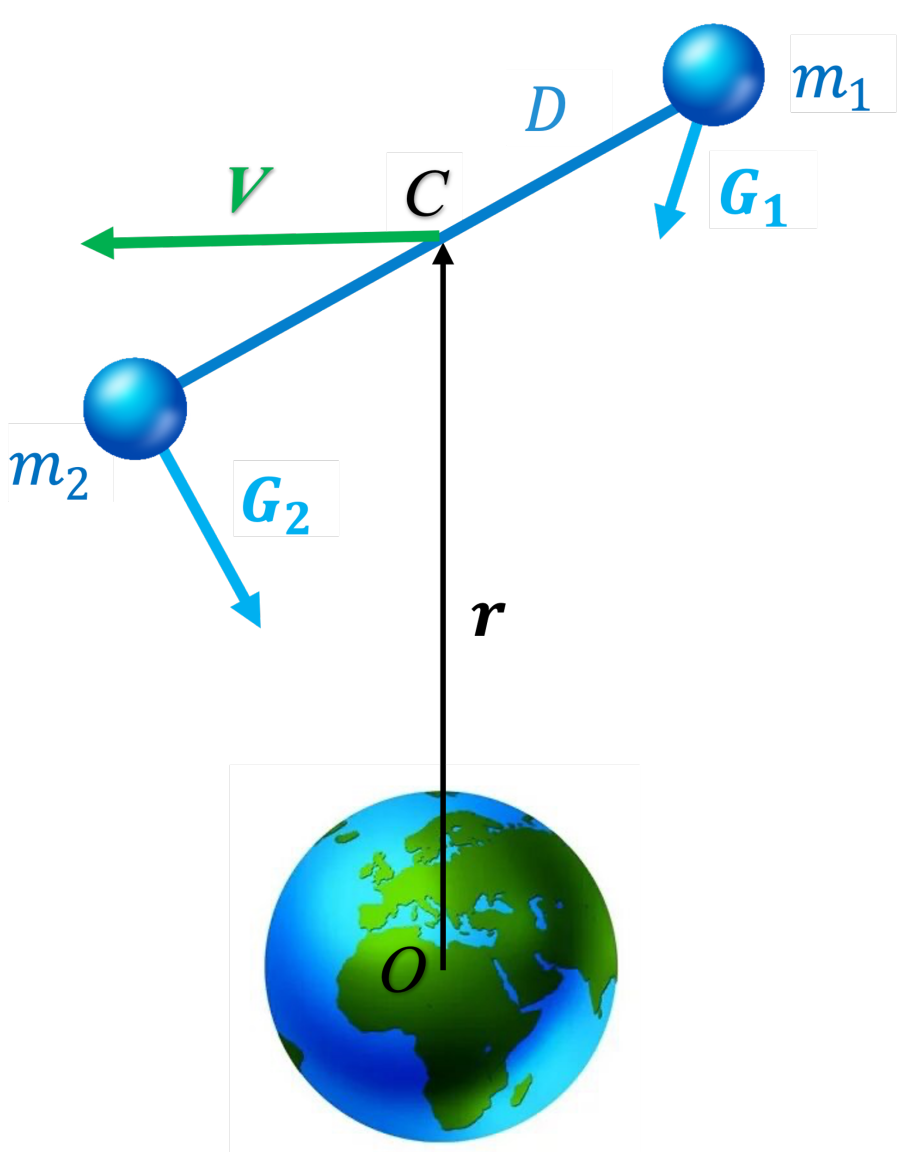

Fig 1. Dumbbell movement in the central gravitational field

The change in the angular momentum of the dumbbell $\boldsymbol{K}_O$ relative to the center $O$ is equal to the main moment of the external forces $\boldsymbol{M}_O$ (angular momentum change theorem)

$$\frac{d\boldsymbol{K}_O}{dt} = \boldsymbol{M}_O \ .$$

The moments of attraction forces $\boldsymbol{G_1}$ and $\boldsymbol{G_2}$ relative to the center $O$ are equal to zero, therefore

$$\boldsymbol{M}_O = 0,$$

and the angular momentum of the dumbbell $\boldsymbol{K}_O$ is a constant value.

$$\boldsymbol{K}_O = \boldsymbol{K}_e + \boldsymbol{K}_i \,;$$

$\boldsymbol{K}_e$ – the vector of the angular momentum of the mass center of the dumbbell $C$, in which the entire mass of the dumbbell is concentrated, relative to the center $O$;

$\boldsymbol{K}_i$ – the vector of the angular momentum of the dumbbell rotation relative to the mass center $C$.

$$\boldsymbol{K}_e = m\,\boldsymbol{r} \times \boldsymbol{V};$$

$m$ – dumbbell mass ($m = m_1 + m_2$);

$\boldsymbol{r}$ – the radius vector of the mass center of the dumbbell to the attractive center *O*;

$\boldsymbol{V}$ – the velocity vector of the mass center *C* of the dumbbell.

$$\boldsymbol{K}_i = J_D\,\boldsymbol{\Omega};$$

$J_D$ – the moment of inertia of the dumbbell in the plane of motion relative to the center *C*, the central axial (binormal) moment of inertia;

$\boldsymbol{\Omega}$ – absolute angular speed of the dumbbell rotation.

The system of attraction forces $\boldsymbol{G}_1$ and $\boldsymbol{G}_2$ for a rigid dumbbell is equivalent to the main vector of the system of forces $\boldsymbol{F_C}$ applied at the center *C*, and the main moment $\boldsymbol{M}_C(\boldsymbol{G_1}, \boldsymbol{G}_2)$ of forces $\boldsymbol{G}_1$ and $\boldsymbol{G}_2$ relative to the center *C*

$$\boldsymbol{F}_C = \boldsymbol{G}_1 + \boldsymbol{G}_2; \qquad (1)$$

$$\boldsymbol{M}_C\,(\boldsymbol{G}_1, \boldsymbol{G}_2) = \boldsymbol{M}_C(\boldsymbol{G}_1) + \boldsymbol{M}_C(\boldsymbol{G}_2). \qquad (2)$$

Let us write equation (1) in projections on the axis of the orbital coordinate system *Cxyz* (Fig. 2):

$$F_{Cx} = G_{1x} + G_{2x}; \qquad (3)$$

$$F_{\boldsymbol{C}x} = \frac{3}{2}\mu_0 \frac{mD^2}{r^4} \frac{\eta}{(1+\eta)^2} \sin 2\varepsilon; \qquad (4)$$

$$F_{Cy} = G_{1y} + G_{2y}; \qquad (5)$$

$$F_{\boldsymbol{C}y} = -\mu_0 \frac{m}{r^2} + 3\mu_0 \frac{mD^2}{r^4} \frac{\eta}{(1+\eta)^2}\left(\frac{1}{2} + \sin^2\varepsilon\right); \qquad (6)$$

ε – the angle between the axis *Cx* of the orbital coordinate system *Cxyz* and the line connecting the end elements of the dumbbell;

$\mu_0 = 3{,}986 \cdot 10^{14}$ m³/s² – geocentric gravitational constant of the Earth;

η – the ratio of the end masses of the dumbbell $m_1$ and $m_2$ or the distances $D_1$ and $D_2$ from the end masses to the center *C* ($D_1 + D_2 = D$)

$$\eta = m_2 / m_1 = D_1 / D_2\,; \qquad (7)$$

Moment $M_C$, seeking to return the dumbbell to a position along the local vertical [10]:

$$M_C = \frac{3}{2}\mu_0 \frac{J_D}{r^3} \sin 2\varepsilon; \qquad (8)$$

A detailed derivation of expressions (4), (6), and (8) is given in the Appendix, as well as in [8].

### 2.2. Thrust Generation

The main vector of the system of forces $\boldsymbol{F_C}$ can also be decomposed into two vectors $\boldsymbol{G}$ and $\boldsymbol{F_T}$ (Fig. 2). Modules of these vectors can be determined from Equations (4) and (6):

$$G = \mu_0 \frac{m}{r^2} + \frac{3}{2}\mu_0 \frac{mD^2}{r^4} \frac{\eta}{(1+\eta)^2};$$

$$F_T = 3\mu_0 \frac{mD^2}{r^4} \frac{\eta}{(1+\eta)^2} \sin\varepsilon;$$

$\boldsymbol{G}$ – gravity force at the mass center $C$, directed along the local vertical towards the center $O$;

$\boldsymbol{F_T}$ – thrust at the mass center $C$, directed along the dumbbell towards the mass $m_1$ at $\sin\varepsilon > 0$ or towards $m_2$ at $\sin\varepsilon < 0$ (Fig. 2).

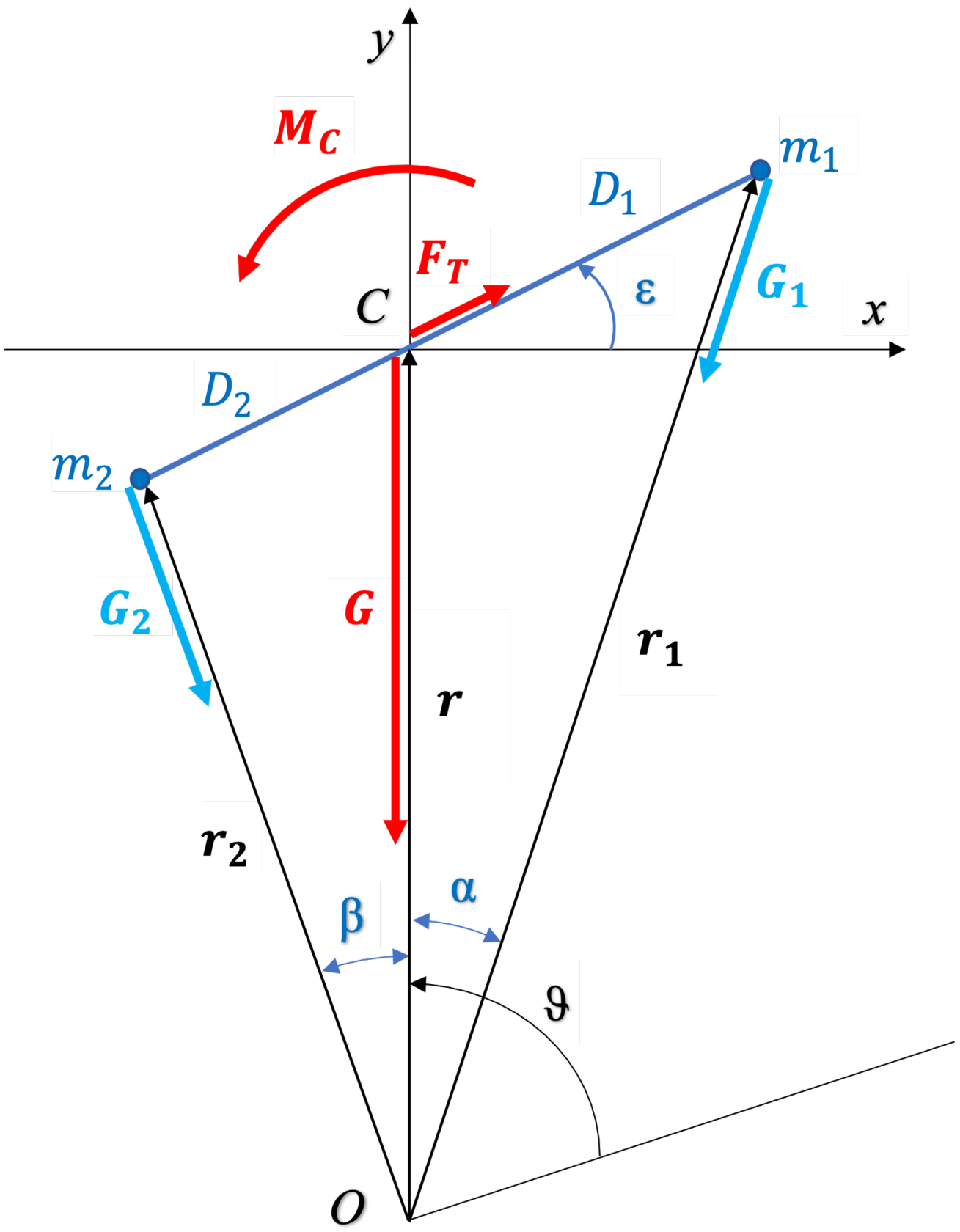

Fig 2. Equivalent systems of forces

The main vector of the system of forces $\boldsymbol{F_C}$ can also be decomposed into two vectors $\boldsymbol{G}$ and $\boldsymbol{F_T}$ (Fig. 2). Modules of these vectors can be determined from Equations (4) and (6):

$$G = \mu_0 \frac{m}{r^2} + \frac{3}{2}\mu_0 \frac{mD^2}{r^4} \frac{\eta}{(1+\eta)^2};$$

$$F_T = 3\mu_0 \frac{mD^2}{r^4} \frac{\eta}{(1+\eta)^2} \sin\varepsilon;$$

$\boldsymbol{G}$ – gravity force at the mass center $C$, directed along the local vertical towards the center $O$;

$\boldsymbol{F_T}$ – thrust at the mass center $C$, directed along the dumbbell towards the mass $m_1$ at $\sin\varepsilon > 0$ or towards $m_2$ at $\sin\varepsilon < 0$ (Fig. 2).

### 2.3. Orbital Control and the Role of the Flywheel

To maintain a given position of the dumbbell at an angle ε, a counterbalancing moment is required, which can be created using a flywheel. The forces of inertia of the flywheel are reduced to a pair of forces with a moment.

$$\boldsymbol{M}_J = -J\dot{\boldsymbol{\omega}};$$

$J$ – flywheel moment of inertia;

$\dot{\boldsymbol{\omega}}$ – angular acceleration of the flywheel rotation.

Thus, the system of equations of motion of the mass center $C$ of the dumbbell in the polar coordinate system $(r, \vartheta)$ with a flywheel of mass $m_J$ that maintains the angle ε (see the Appendix):

$$\begin{gathered} \ddot{r} - \dot{\vartheta}^2 r = -\frac{\mu_0}{r^2} + 3\mu_0 \frac{D^2}{r^4} \frac{\eta}{(1+\eta)^2} \left(\frac{1}{2} + \sin^2\varepsilon\right) \frac{m}{(m+m_J)}; \\ r\ddot{\vartheta} + 2\dot{r}\dot{\vartheta} = -\frac{3}{2}\mu_0 \frac{D^2}{r^4} \frac{\eta}{(1+\eta)^2} \frac{m}{(m+m_J)} \sin 2\varepsilon\,. \end{gathered} \tag{9}$$

On the basis of the system of equations (9), mathematical modeling of the change in the radius $\Delta r = r_0 - r$ on two orbits was carried out under the initial conditions: $r_0 = 6\,675$ km, $\dot{\vartheta}_0 = 0.001157689\ \mathrm{s}^{-1}$, $D = 100$ km, $\eta = 1$, $m = m_J$ (Fig. 3). In the case of $\varepsilon = 3\pi/4$, $r$ increased by 7 km per one orbit, and in the case of $\varepsilon = \pi/4$, it decreased by 7 km.

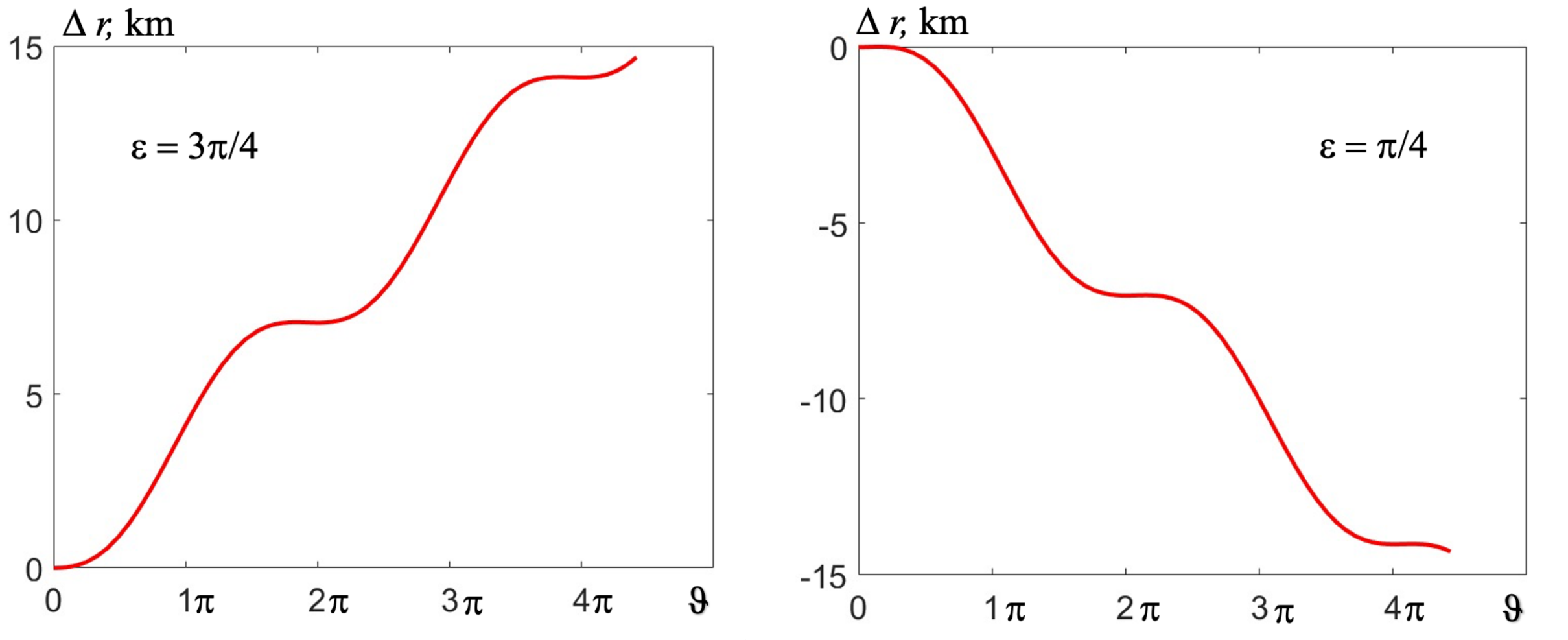

Fig 3. Changing the radial displacement of the dumbbell mass center

As a result, spinning the flywheel to a certain angular velocity $\boldsymbol{\omega}$, the angular momentum $\boldsymbol{K}_i$ changes, and, consequently, the angular momentum $\boldsymbol{K}_e$. The limitation on the maximum change in $\boldsymbol{K}_e$ is due to the limiting angular velocity of the flywheel rotation.

Fig. 4 shows a diagram of the radial movement of the mass center of the dumbbell $C$. By changing the direction of the flywheels rotation, the movement of the system can be carried out down (Fig. 4 a) and up (Fig. 4 b). The travel range is limited by the maximum angular speed of the flywheel rotation.

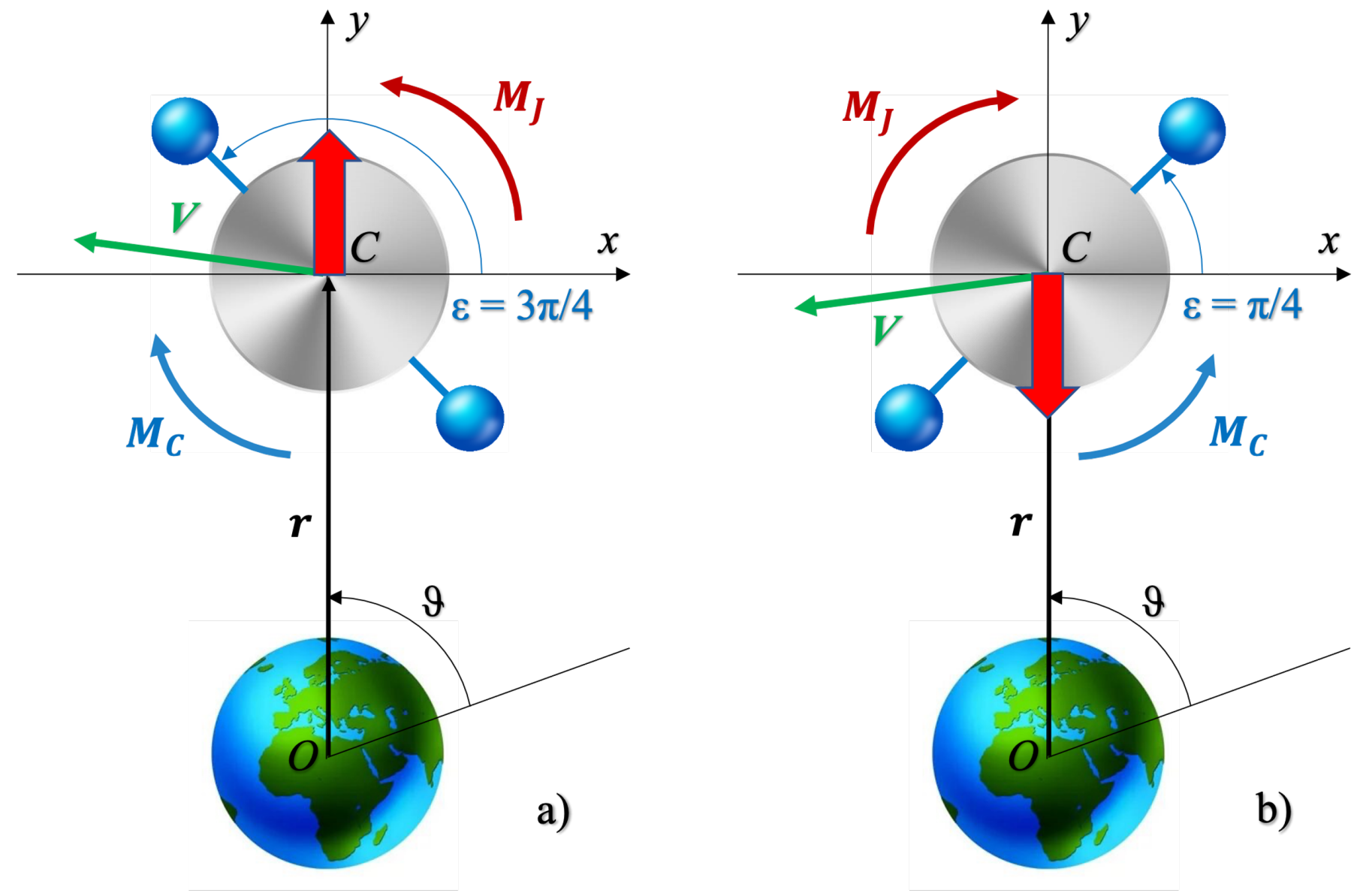

Fig. 4. Change in the radial displacement of the mass center of the dumbbell

By arranging flywheels at different orbital heights within the same plane, it is possible to implement a scheme for the movement of counter-flow cargo streams without fuel consumption (see Fig. 5). The payload, represented by a dumbbell, moves from one flywheel to another. Controlling the angle of inclination of the dumbbell allows for a 'soft encounter' (where the relative speed reaches zero) with the next flywheel in orbit. To spin the flywheels, only electrical energy from power sources (such as solar panels) is required. When moving upwards, the flywheels are spun with angular acceleration in one direction, while during downward movement, they are spun in the opposite direction. This process unloads the flywheels and conserves electrical energy for their acceleration. The difference in heights between the flywheels is limited by the maximum displacement of the dumbbell-flywheel system.

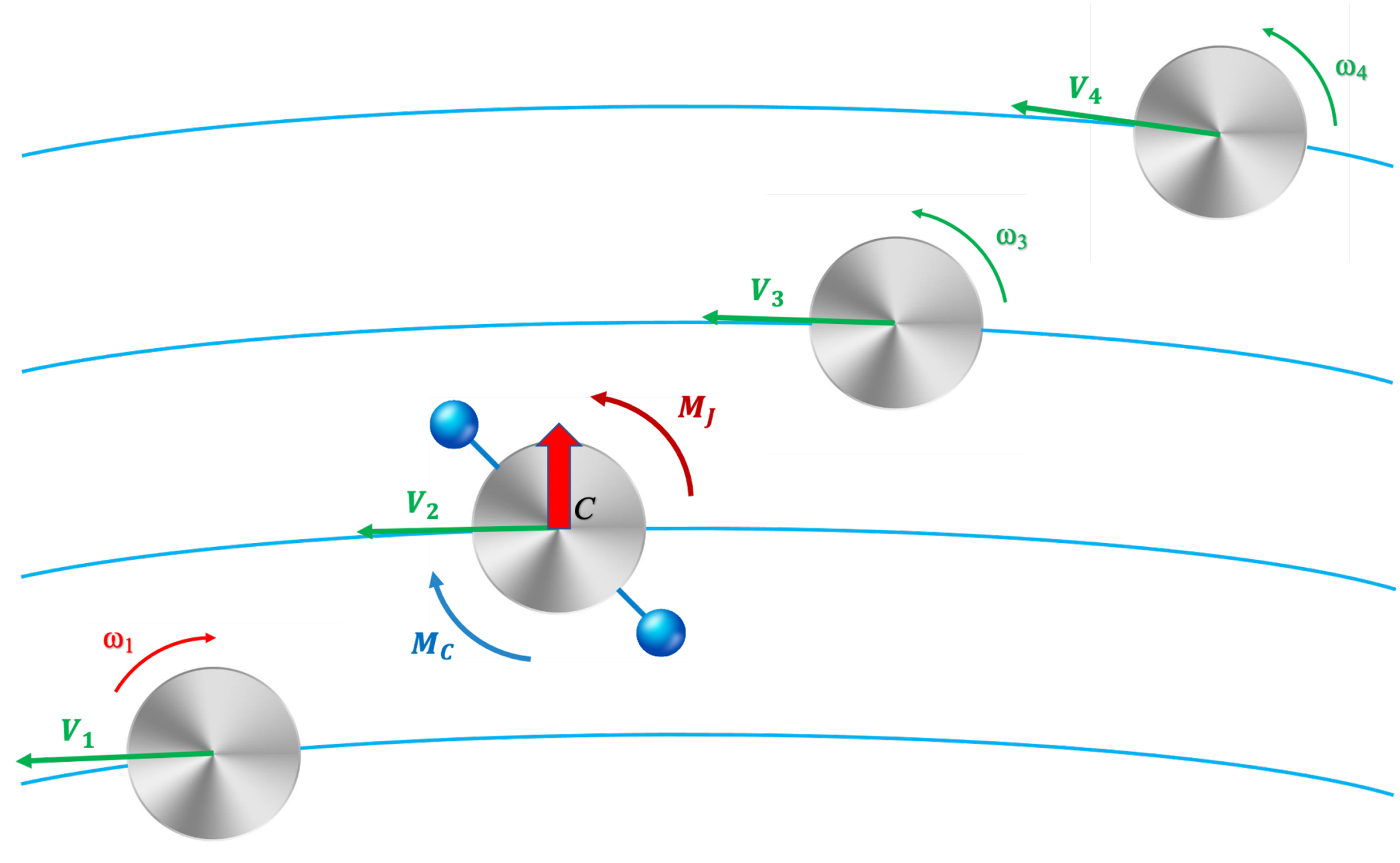

Fig. 5. Scheme of movement in the radial direction

However, the technical implementation and efficiency of orbital maneuvers of this scheme [13] is inferior to maneuvers for the exchange of kinetic energy with the use of tether systems technologies [14, 15].

The fact of the relationship between rotational motion around the mass center and radial motion is observed in nature. Every year the Moon moves away from the Earth by 3.8 cm, while the Earth slows down its angular velocity of rotation [16].

Thus, the relationship between the rotational motion of the body relative to the mass center and the radial motion of the body is shown. It should be noted that there is no violation of the conservation law of the mass center position. The center of the gravitational field $O$ (the mass center of a closed system, and more strictly - the mass center of the Earth-dumbbell system), as well as the mass center of the Earth-Moon system, does not change its position. Only the position of the bodies relative to the common mass center changes.

### 2.4. Conclusions and Problem Statement

The "dumbbell-flywheel" system demonstrates the fundamental feasibility of generating thrust and controlling orbital motion through redistribution of internal kinetic momentum in a non-uniform gravitational field.

Quantitative assessment reveals that the macroscopic manifestation of this effect is negligible for practical propulsion applications due to its strong $D^2/r^4$ dependence.

A logical progression of this principle involves its extrapolation to the quantum level, where elementary particle spins can function as elemental "flywheels." This approach enables achieving orders of magnitude higher energy efficiency compared to macroscopic analogs, justifying the transition to investigating the quantum hypothesis.

## 3. Extrapolation of the Principle to the Quantum Level

Classical analysis has revealed both a fundamental possibility and a fundamental limitation for macroscopic systems. This section substantiates the feasibility of transferring the considered principle to the quantum level, where the spin of elementary particles serves as the carrier of angular momentum. The aim of this section is to demonstrate that such a transfer paves the way for creating a system with potentially orders of magnitude higher energy efficiency.

### 3.1. Spin as a Quantum Analog of a Flywheel

In quantum mechanics, spin represents an intrinsic, indestructible angular momentum of an elementary particle, unrelated to its motion in space [17]. In its physical dimension, spin is identical to the angular momentum of a flywheel. This allows for the hypothesis that controlling the spin states of a particle ensemble (their orientation or flux) can play a role analogous to changing the angular momentum of a rotating flywheel in a classical system. A change in the total spin of a system of bodies must, in accordance with general conservation laws, be compensated by a change in its orbital motion (Fig. 4).

It is known from quantum mechanics [17] that elementary particles have spin (intrinsic angular momentum), which has a quantum nature and it is not associated with the movement of the particle as a whole.

Let use elementary particles as flywheels (Fig. 6). Change in the angular momentum $\boldsymbol{K}$ of the moved object due to the radiation of n elementary particles

$$\Delta \boldsymbol{K} = -n\,\boldsymbol{s}\frac{h}{2\pi}\ ;$$

$\boldsymbol{s}$ – the spin vector of an elementary particle;

$h$ – Planck's constant ($h = 6.626070040 \cdot 10^{-34}$ J · s).

In this case, it should be expected that the internal angular momentum $\boldsymbol{K}_i$ of the moved object of mass $m$ will not change (otherwise, we get the unwinding of the object, which can be turned into radial motion using a dumbbell):

$$\Delta \boldsymbol{K} = \Delta(m\,\boldsymbol{r} \times \mathbf{V}) = -n\,\boldsymbol{s}\frac{h}{2\pi}\ ;$$

In a fairly short period of time

$$\Delta(m\,\boldsymbol{r} \times \mathbf{V}) \cong m\,\boldsymbol{r} \times \Delta \boldsymbol{V}_{\boldsymbol{K}}\ ,$$

$\Delta \boldsymbol{V}_{\boldsymbol{K}}$ – the vector of change in the velocity of an object of mass $m$, in the case of a change in its angular momentum $\boldsymbol{K}$ due to the radiation of $n$ elementary particles.

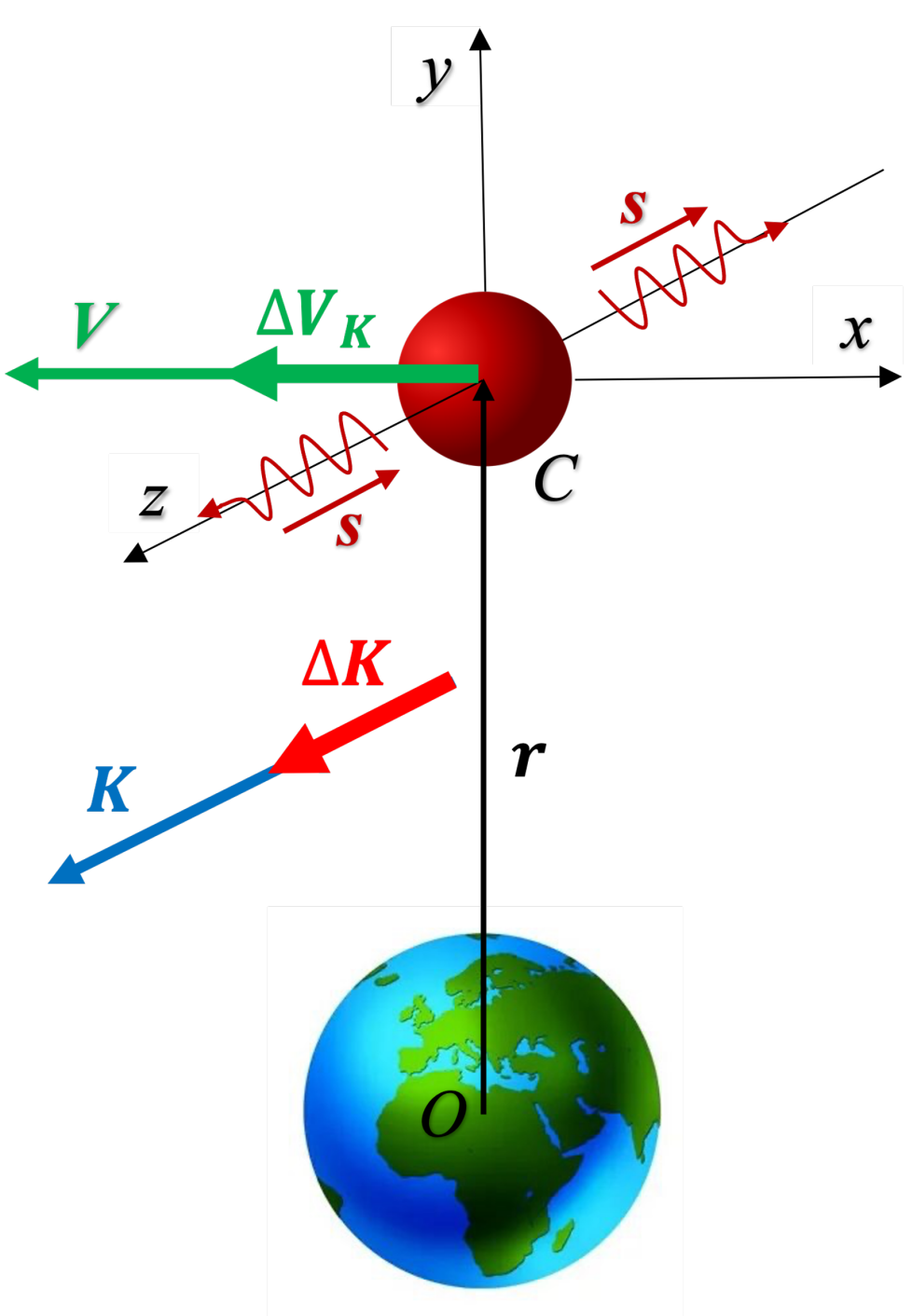

Fig. 6. Motion based on the use of the elementary particles spin

In scalar form

$$m\, r\, \Delta V_K \sin(\angle(\boldsymbol{r}, \Delta \boldsymbol{V_K})) = n \frac{s\, h}{2\pi}$$

In case of $\sin(\angle(\boldsymbol{r}, \Delta \boldsymbol{V_K})) = 1$

$$m\, \Delta V_K = n \frac{s\, h}{2\, \pi\, r}. \qquad (10)$$

**3.2. Comparative Analysis of Energy Efficiency**

Let us consider the latter expression from the perspective of energy costs for motion based on the application of changes in kinetic moment and momentum (reactive motion).

To evaluate the effectiveness of the proposed principle, a comparative analysis is conducted with an ideal photon engine – the most efficient type of rocket engine, where the exhaust velocity equals the speed of light.

Photon Engine (Reactive Principle). The change in momentum of a body of mass *m* by an amount $\Delta V_J$ due to the emission of *n* photons with wavelength λ:

$$m\, \Delta V_J = n\, \frac{h}{\lambda}, \qquad (11)$$

$\Delta V_J$ – the vector of change in the speed of an object of mass *m* in the case of jet propulsion due to the radiation of *n* photons with a wavelength λ.

In this case, energy costs for movement:

$$\Delta E_J = n\, \frac{h\, c}{\lambda}, \qquad (12)$$

where *c* – the speed of light.

From expressions (10) and (11) it follows that for $\lambda > 2\pi r/s$, to change the velocity of an object in a central field at a distance *R* from the center of attraction, it is energetically more advantageous to use the angular momentum of an elementary particle in comparison with its momentum (jet motion). In this case, the radiation of low-energy particles should be carried out in the direction perpendicular to the plane of motion (Fig. 6). The obtained results theoretically prove the feasibility and energy efficiency of implementing the thrust creation hypothesis based on kinetic moment alteration for developing transport vehicles utilizing new physical principles.

Let us evaluate the practical feasibility of the hypothesis implementation. Currently, the generally accepted classical theory of gravity remains Albert Einstein's general theory of relativity (GTR). Within its framework, particles are considered as point-like objects moving along deterministic trajectories in curved spacetime. However, since the gravitational field in GTR is described by a tensor field (the spacetime metric), its quantum carrier, by analogy with other fundamental interactions, should also be a tensor particle. Within the formalism of quantum field theory (QFT), this requirement is satisfied by a massless boson with spin 2. Thus, the consistent application of quantization principles leads to the widely accepted hypothesis that the gravitational field at the quantum level is described by the graviton – precisely such a particle.

**3.3. Order-of-Magnitude Estimate of the Effect**

For illustration, we will perform an approximate estimation. As an example, let us consider an elementary particle of the "graviton" type. The Compton graviton wavelength $\lambda_g > 1 \cdot 10^{16}$ m [18], which is much larger than the Earth's radius and the distance from the Earth to the Sun. Thus, if gravitons are used for motion, then using their spin (angular momentum) (10) is a billion times more profitable (12) than using them in jet motion near the Earth's surface (11). The spin vector ***s*** (direction of radiation) is directed perpendicular to the plane of motion of the object.

Let's estimate the acceleration that the object receives:

$$a = \frac{\Delta V}{\Delta t} = \frac{s\,h}{2\pi r m \Delta t}.$$

The possibility of controlling quantum processes with an accuracy of up to three attoseconds has been proven ($\Delta t = 3 \cdot 10^{-18}$ s) [19]. Spin graviton $s = 2$. Neutron (proton) mass $1.675 \cdot 10^{-27}$ kg ($m = 1.675 \cdot 10^{-27}$ kg). $r = 6.371 \cdot 10^{6}$ m. Then acceleration will act on each neutron (proton) $a = 6{,}600$ m/s$^2$.

Thus, elusive, massless gravitons are capable of setting in motion neutrons and protons, which are gigantic in comparison. Note that the presented methodology for calculating acceleration is valid for all elementary particles possessing spin, including the carrier particles of fundamental interactions.

This estimate, although speculative, demonstrates that the hypothetical mechanism is capable of generating significant effects at the microlevel. The task of experimental physics is to investigate the possibility of generating and detecting directed fluxes of such particles by a macroscopic object.

Taking into account nature's tendency to achieve object motion with minimal expenditure (the principle of least action), a simple logical conclusion suggests itself regarding "virtual photons" – the carriers of electromagnetic interaction: their action effect may be mediated by particles of the "graviton" type.

An intriguing question arises: are such accelerations feasible for macroscopic objects? If we could induce all their atoms to simultaneously absorb/emit such low-energy particles, we would achieve motion without internal deformation – that is, motion without overload. Practical implementation of this hypothesis requires generating directed fluxes of low-energy particles.

**3.4. Conclusions and Problem Statement for Experimental Verification**

Theoretical analysis allows us to formulate the following propositions:

Transferring the principle of angular momentum control to the quantum level, where the spin of elementary particles acts as a flywheel, theoretically enables the creation of propulsion systems with radically higher energy efficiency compared to reactive methods.

Realizing this advantage requires fulfilling a fundamental condition: an accelerating body must asymmetrically emit or absorb fluxes of low-energy spin-carrying particles in a direction perpendicular to its velocity vector.

The primary experimental objective is the detection and registration of the existence of such spin-polarized fluxes associated with the accelerated motion of macroscopic bodies.

## 4. Mechanistic Model and Pilot Experiments

This section serves as a bridge between the theoretical foundation and the program of future research. Its purpose is to formulate a working hypothesis that provides a mechanistic explanation for the connection between accelerated motion and spin fluxes, as well as to present the results of pilot experiments that, while not constituting proof, indicate the presence of anomalies requiring further investigation.

### 4.1. Hypothesis Formulation

What is the difference between a material particle and the same one, but moving 100 m/s faster? The modern theories accepted by the scientific community correspond to: the difference in kinetic energy $\Delta E$, and, based on the principle of equivalence of mass and energy, the change in relativistic mass $\Delta m = \Delta E / c^2$.

The aforementioned fact, the adherence to the laws of conservation of energy, momentum, and angular momentum for a closed system in quantum mechanics, as well as the provided example with gravitons in the scheme of radial object motion (Fig. 6), provide grounds for formulating a mechanistic theory of interaction carriers and proposing the hypothesis of the spin-density nature of interactions and inertia.

Core Proposition. Space-time represents a dynamic medium composed of low-energy interaction carrier particles (conventionally termed "gravitons"), possessing spin and existing in two fundamental states: free (perturbation field) and bound (constituting relativistic mass). The geometry of space-time and gravitational interaction are macroscopic manifestations of the density, spin polarization, and state (free/bound) of this medium.

Implication for Accelerated Motion. Accelerated motion of a macroscopic body is accompanied by transitions of low-energy particles between free and bound states. This process generates two interconnected "wakes":

1. Density Oscillation Wake – a wave disturbance in the distribution of free particles, caused by transitions and propagating spherically.

2. Spin Wake – an asymmetric spin polarization of low-energy particles in the plane perpendicular to the instantaneous velocity vector.

While the density wave propagates spherically, the asymmetry in particle density and spin state critical for the emergence of inertial force forms predominantly in the plane perpendicular to the instantaneous velocity vector. It is precisely this local anisotropy that creates the force of inertial resistance to acceleration.

The hypothesis is based on the following postulates and mechanisms:

1. Ontological Basis.

Carrier Medium: Space is filled with real low-energy carrier particles. Their concentration and spin state determine the local "curvature" and properties of space-time.

Two States: Particles can exist in a free state (forming the field) or in a bound state with matter (constituting its mass). The total particle density (free + bound) in a closed system is constant.

2. Mechanism of Gravity and Inertia.

Gravity: Caused by a pressure gradient arising from the difference in the density of free particles. In the vicinity of a massive body, the density of free particles decreases as some of them transition to a bound state. This density difference creates a force directed towards the body – gravitational attraction.

Inertia (Causality): A change in the state of motion of a body (acceleration) is the cause that induces a local disruption of the dynamic equilibrium between the absorption (transition to bound state) and emission (transition to free state) of carrier particles.

3. Directionality of Spin Fluxes.

During body acceleration, the flux of absorbed (bound) particles exceeds the flux of emitted (released) particles; during deceleration, the opposite occurs – emission exceeds absorption. In steady, uniform motion, the fluxes are balanced. This asymmetry in the number of transitions is accompanied by their spatial and spin anisotropy.

4. Connection with Quantum Phenomena.

Wave-Particle Duality: The motion of a material particle represents the motion of a stable wave packet, formed and maintained by its interaction with the polarized medium. This explains interference phenomena, including the double-slit experiment.

"Pilot Wave": The wake of oscillations and spin polarization created by the accelerated motion of an elementary particle (e.g., an electron) acts as a guiding "pilot wave" in the sense of the de Broglie-Bohm theory, determining the probabilistic nature of motion.

Role of the Observer: Active measurement (observation), interacting with the medium, destroys or significantly modifies its fine spin structure ("wake" and "pilot wave"), leading to the collapse of the wave packet.

5. Conservation Laws.

This mechanism ensures the fulfillment of the laws of conservation of energy, momentum, and angular momentum for the closed system "body + field of carrier particles," as all processes reduce to the redistribution and transformation of the state of particles within the system.

Conclusion. The proposed approach provides a mechanistic interpretation that seeks to connect classical motion, the geometry of General Relativity, and quantum phenomena, postulating that the primary entities are hypothetical low-energy particles whose distribution and state directly shape physical reality.

According to this hypothesis, its foundation lies in the emission and absorption of low-energy particles. To estimate the momentum of emitted/absorbed elementary particles, we write equation (10) in the following form:

$$m\,\Delta V = n\frac{s\,h}{2\pi\,\rho}, \qquad (13)$$

ρ – the average radius of space curvature in the vicinity of material quantum particles of the body, due to the forces of gravitational attraction of the universe. The use of this scalar parameter in (13) is due to the relationship between the rotational motion of the body relative to the mass center and the radial motion of the body. The specific value of the introduced parameter for the subsequent estimation of the momentum of emitted/absorbed elementary particles is not of fundamental importance in this work, but, of course, is of interest for further research.

An impulse $I_r$ of radiation/absorption

$$I_r = n\frac{s\,h}{2\pi\,\rho} \qquad (14)$$

appears as a result of a change in the speed of a body and can be transmitted to other bodies. As a consequence, the law of conservation of momentum takes on a broader interpretation: the momentum of the system $I$ and the radiation/absorption momentum of elementary particles $I_r$ for a closed system is a constant value

$$I + I_r = const\,,$$

regardless of the type of interaction of the bodies of the system (elastic or inelastic impact).

To confirm the hypothesis put forward, it is advisable to consider a number of examples from different areas of physics and conduct experiments to search for the pulse of radiation/absorption of elementary particles $I_r$.

**4.2. Indirect Arguments and Analogies**

The well-known electron diffraction experiment (Fig. 7) demonstrates the emergence of a momentum component perpendicular to the slit after electrons pass through it. From the perspective of the proposed hypothesis, this effect can be explained by the emission and absorption of low-energy carrier particles. It is noteworthy that the quantum uncertainty of elementary particles may result from their constant interaction with this dynamic medium.

Transitions of low-energy particles between free and bound states are accompanied by density oscillations and changes in the spin polarization of spacetime. This process creates a wave structure that fulfills the function of a "pilot wave" in the de Broglie-Bohm theory [20]. The emitted waves propagate along the electron's trajectory, forming a wavefront in the perpendicular plane. Subsequent interaction of the electron with the polarized medium determines its motion in the form of a wave packet and explains the emergence of interference patterns, including the classic double-slit experiment.

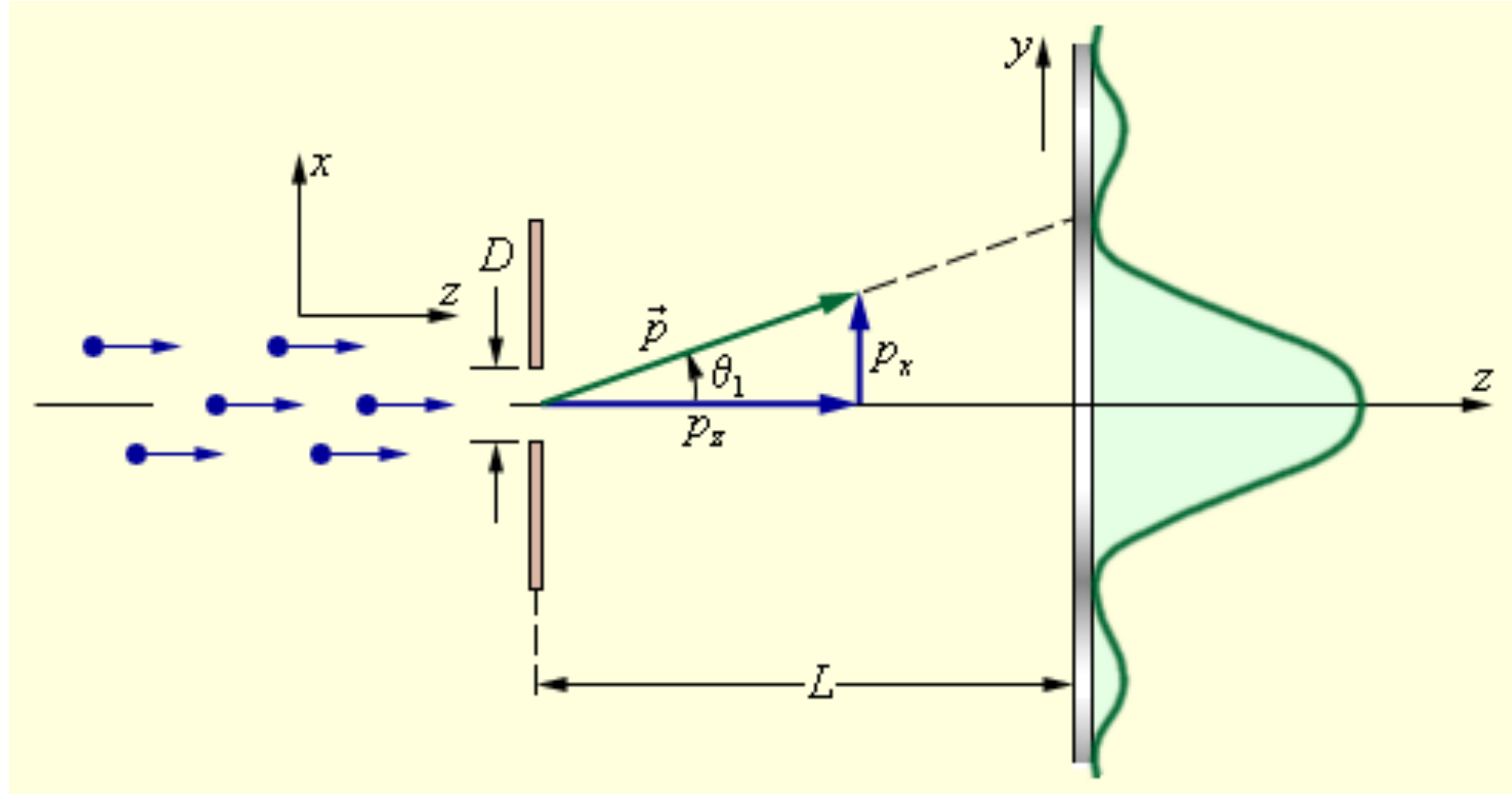

Fig. 7. Diffraction of electrons by the slit

The observer effect in quantum phenomena receives a natural explanation within this hypothesis through the disruption of the medium's fine spin structure. The hypothesis predicts a similar influence on electron fluxes by any accelerated material body, including freely falling objects, opening prospects for experimental verification.

Figure 8 shows frames from high-speed video of a falling droplet [21]. Following a central inelastic impact, the liquid spreads predominantly in the plane perpendicular to the droplet's original direction of motion, corresponding to a transformation of the vertical momentum component into horizontal momentum.

From the perspective of the proposed hypothesis, this phenomenon is explained by the formation of two interconnected structures—the density oscillation wake and the spin wake. The vertical momentum component is converted into emission of low-energy particles in the horizontal plane, generating a spherically propagating density oscillation wake. Simultaneously, a spin wake forms—an asymmetric polarization of particles in the plane perpendicular to the instantaneous velocity vector.

According to the proposed mechanism, water molecules acquire momentum in the horizontal plane, perpendicular to the direction of the density wave's propagation, corresponding to momentum transformation via a "90° + 90°" scheme. The spin wake plays a critical role in this process, providing directional action on the liquid molecules. It is precisely the local anisotropy in the spin state of the particles that induces the liquid's preferential motion in the horizontal plane.

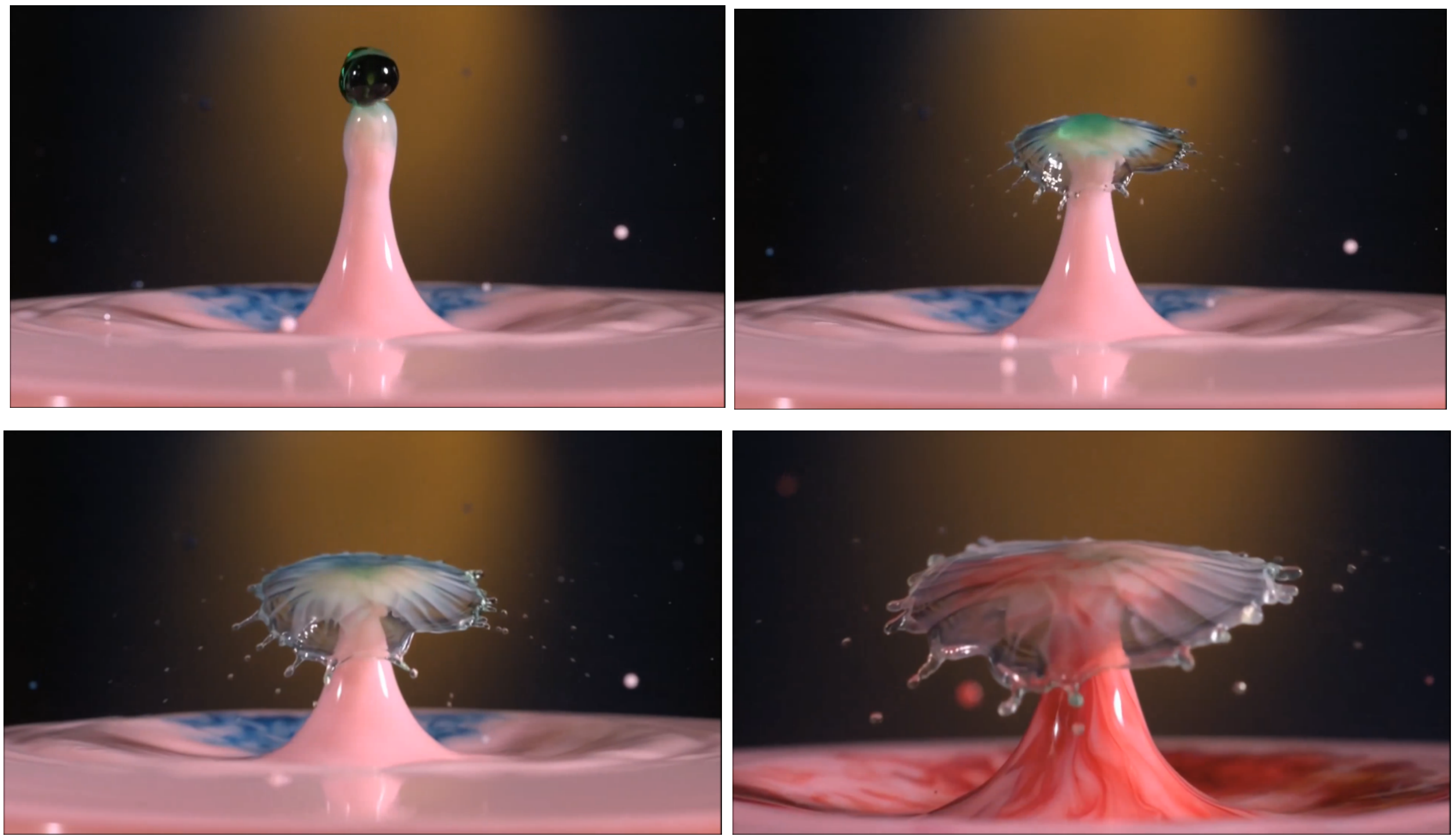

Fig. 8. Central inelastic impact of a liquid drop

A similar mechanism is observed in the case of a central elastic impact, where interconnected density oscillation and spin polarization wakes are also formed. The impact momentum is converted into emission of low-energy particles in the plane perpendicular to the impact direction. Subsequent absorption of this radiation by the molecules of the elastic body in the presence of the spin wake leads to a reversal of their direction of motion by 180°, implementing the same "90° + 90°" scheme.

The universality of this mechanism is confirmed by its manifestation in electromagnetic processes. The electric and magnetic field strength vectors, mutually perpendicular to the wave propagation direction, are interpreted within the hypothesis as manifestations of the density oscillation wake and spin wake, respectively.

Thus, diverse physical phenomena – from elementary particle diffraction to hydrodynamic processes and electromagnetic wave propagation – demonstrate the universal nature of momentum transformation via the "90° + 90°" scheme through the emission/absorption of low-energy particles.

Experimental verification of the hypothesis requires research under vacuum conditions to eliminate the influence of the air medium and directly observe the manifestations of the described interaction mechanisms.

### 4.3. American Vacuum Drop Experiment

On the Internet there is a popular experiment with gravity, which was conducted by physicist Brian Cox in a large vacuum chamber "Space Power Facility" NASA in the US state of Ohio [22]. Fig. 9 shows the time-lapse footage of the fall of a lead ball and a feather in a vacuum. Let us pay attention to the movement of villi against the center of the ball.

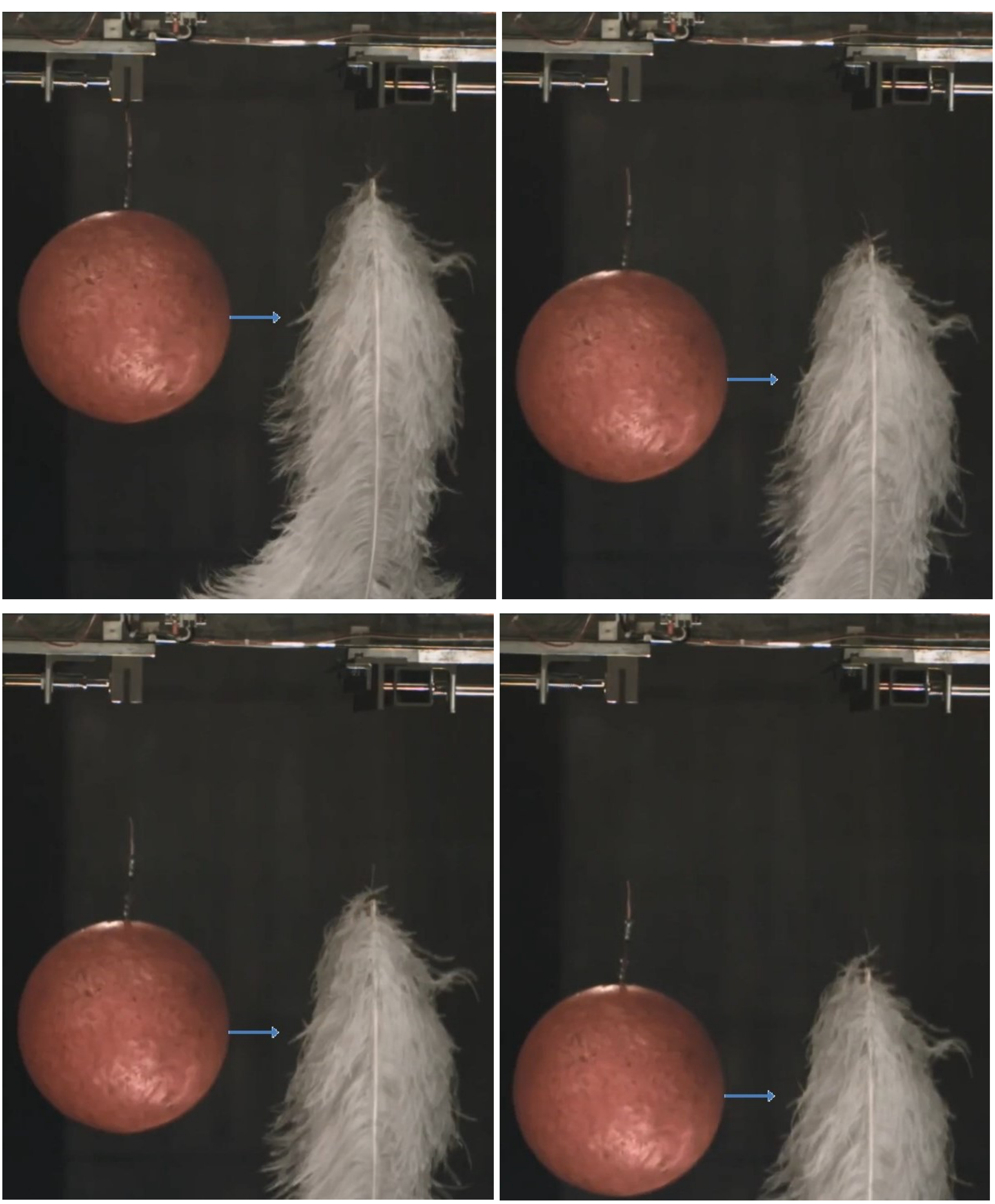

Fig. 9. Footage of a lead ball and feather falling in vacuum

After the simultaneous release of the ball and feather from the attachment in the first moments of falling on the video recording of the experiment [22], the movement of the feather villi is observed, due to their elastic properties during the transition from the suspended state of the feather to weightlessness (free fall). In subsequent moments, the movement of the feather villi facing the center of the ball differs from the general movement of the remaining villi. The nature of the movement of the villi, which the arrow points to (Fig. 9), confirms the assumption about the presence of absorption by the ball of elementary particles in the plane perpendicular to its movement (Fig. 10). In fig. 10 arrows in the plane indicate the direction of motion of elementary particles towards the mass center of the ball. Under the action of radiation according to the “90° + 90°” scheme, the ends of the villi in the horizontal plane of the center of mass of the ball rise (the law of conservation of momentum: a massive ball picks up speed, absorbs radiation – the concentration of radiation in the vicinity of light villi changes according to the principle of inverse reaction).

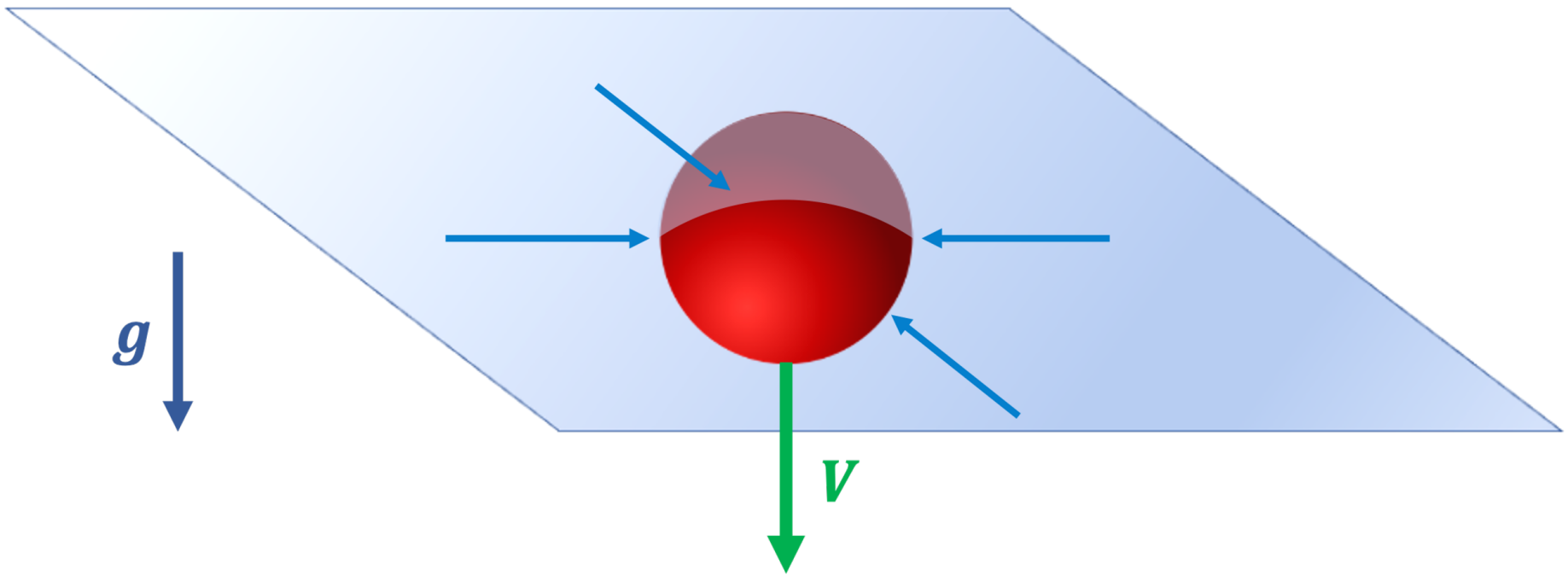

Fig. 10. Directions of motion of elementary particles in the case of accelerated motion

The process of ball rebound from the damper is also under the study. In this case, when the ball moves up, the presented frames of slow motion (Fig. 11) coincides with the assumption that the ball emits elementary particles in the plane perpendicular to its motion. Under the action of radiation according to the “90° + 90°” scheme, the villi in the horizontal plane of the center of mass of the ball rise after the impact and rebound of the ball (the law of conservation of momentum: the ball loses speed, radiates - the concentration of radiation in the vicinity of the villi changes, the villi rise according to the principle of reverse reaction).

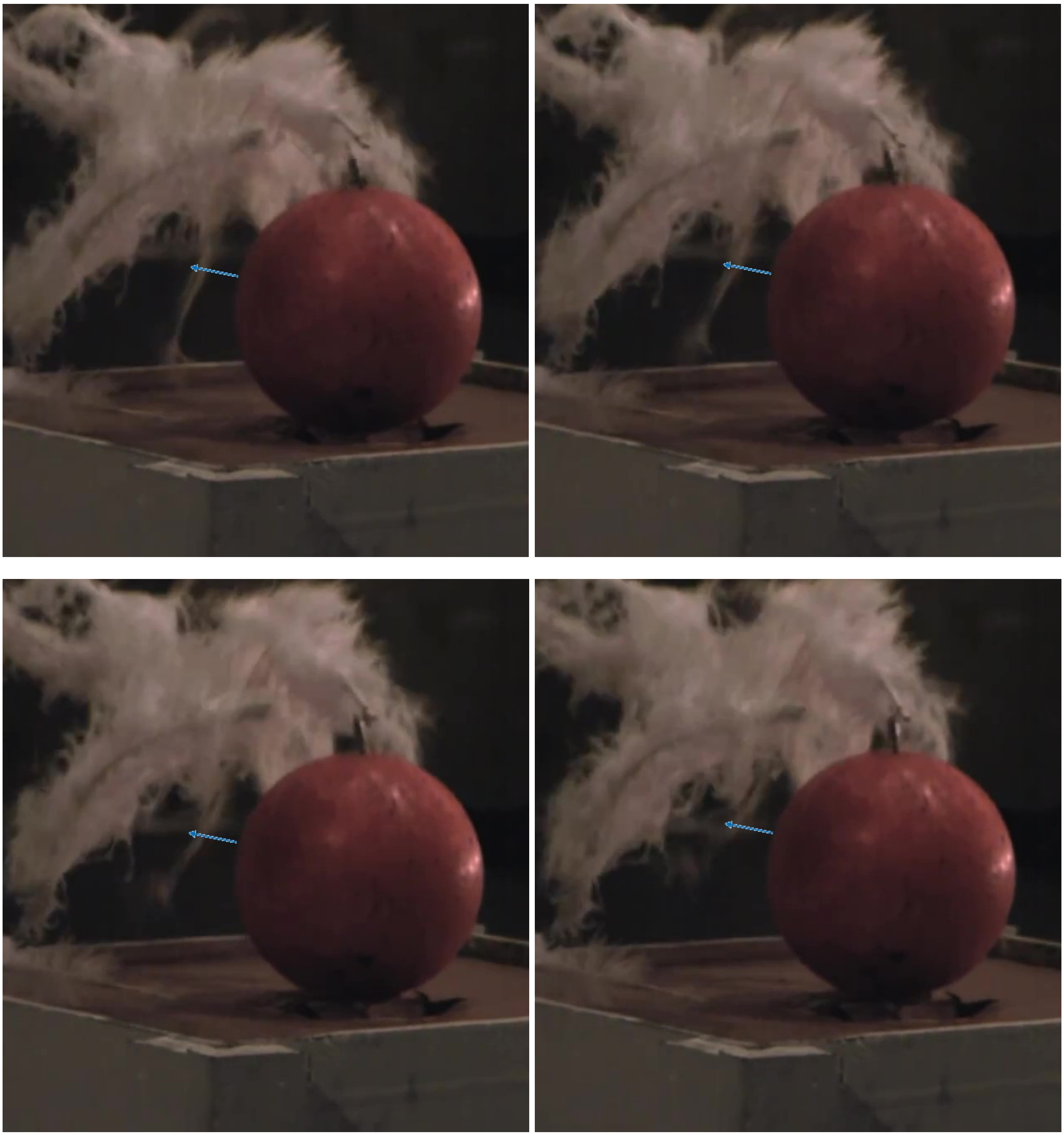

Fig. 11. Frames of the rebound of the ball and feathers in vacuum

The directions of motion of elementary particles from the mass center of the sphere are indicated by arrows in the plane in Fig. 12.

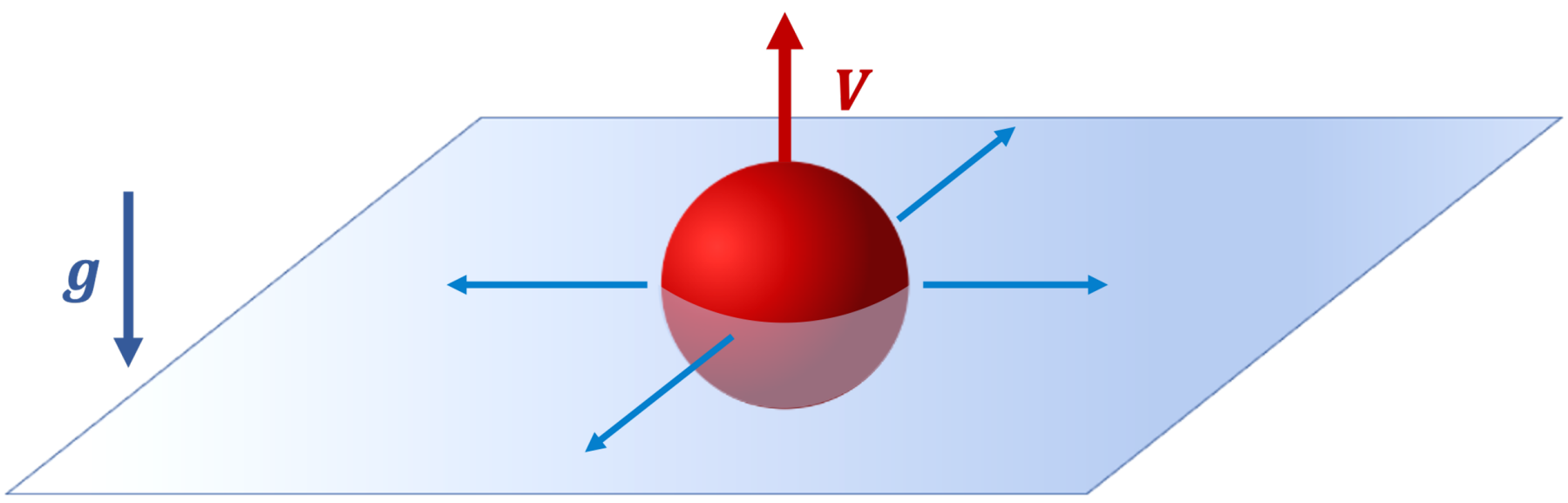

Fig. 12. Directions of movement of elementary particles in the case of slow motion

In the above experiment, feathers play the role of a detector that records the flow of passing elementary particles. The number of these $n_d$ particles can be estimated based on the equation (14):

$$n_d = m\,\Delta V \frac{2\pi\,\rho}{s\,h} \cdot \frac{d\,V\,\Delta t}{2\pi\,l\;V\,\Delta t} = m\,\Delta V \frac{\rho\,d}{s\,h\,l}\ ; \tag{15}$$

$V$ – the speed of the body (ball) relative to the detector (Fig. 13);

$\Delta t$ – the time interval during which the speed of the body (ball) changes by $\Delta V$ with respect to the detector;

$d$ – width of the detector (villi) in the plane of radiation/absorption by the body (ball) of elementary particles (Fig. 13);

$l$ – the distance from the detector (villi) to the mass center of the ball (Fig. 13).

Thus, the area $A = d \cdot V \cdot \Delta t$ crosses $n_d$ elementary particles in time $\Delta t$.

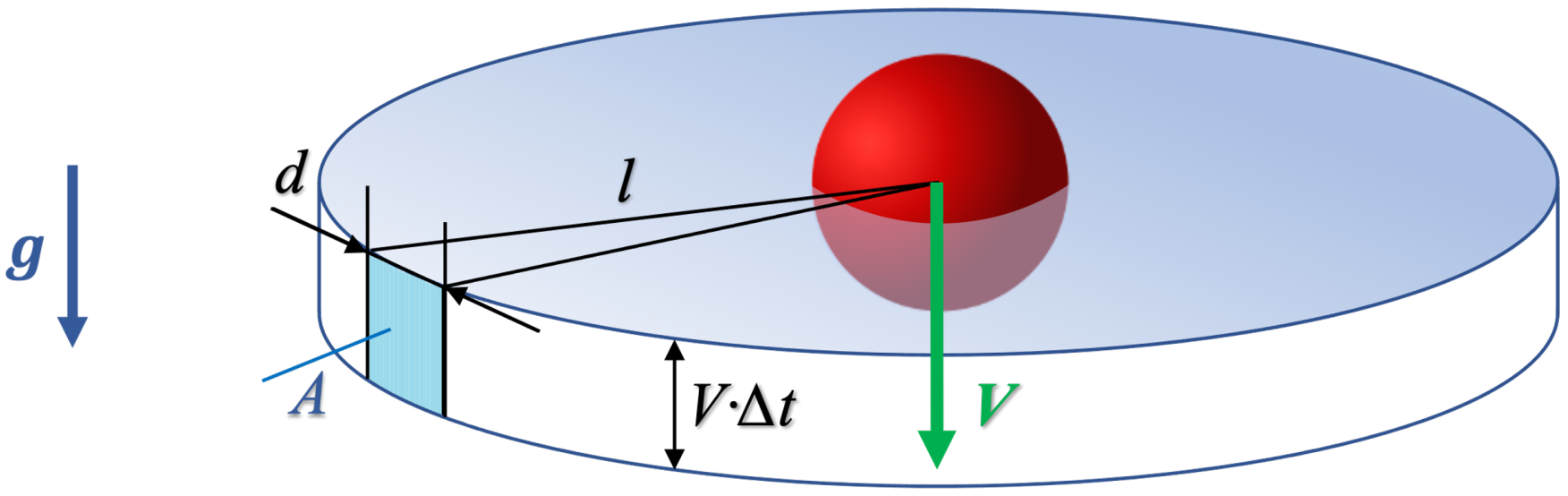

Fig. 13. Scheme for calculating the particle flux through detector A

In the case of a free fall of a body (ball) with acceleration $g$, the dependence of the change in velocity $\Delta V$ on the change in the height of the ball $\Delta H$ is determined as follows:

$$\Delta H = V_0 \cdot \Delta t + \frac{g \cdot \Delta t^2}{2}, \qquad \Delta V = g \cdot \Delta t\,,$$

where $V_0$ – the initial speed of the body – (ball);

$$\Delta H = \frac{V_0 \cdot \Delta V}{g} + \frac{\Delta V^2}{2 \cdot g}\ ;$$

$$\Delta V = \sqrt{V_0^2 + 2g\,\Delta H} - V_0\,. \tag{16}$$

Then, taking into account (15) and (16), the number of particles $n_d$ passing through the detector with height $\Delta H$ is

$$n_d = m\,\frac{\rho\,d}{s\,h\,l}\left(\sqrt{V_0^2 + 2g\,\Delta H} - V_0\right). \tag{17}$$

The function $n_d(V_0)$, defined by expression (17), for positive values of $\Delta H$ has a negative derivative: $n_d'(V_0) < 0$,, therefore, the function$n_d(V_0)$ is decreasing, and its maximum value is attained at $V_0 = 0$:

$$n_{d\,max} = m\,\frac{\rho\,d}{s\,h\,l}\,\sqrt{2g\,\Delta H}. \tag{18}$$

Let us estimate the Compton wavelength of an elementary particle based on the principle of equivalence of mass and energy.

When the speed of a body changes, its relativistic mass changes:

$$\Delta m = m_{rel} - m = m\left(1 \Big/ \sqrt{1 - \frac{\Delta V^2}{c^2}} - 1\right) = m\frac{\Delta V^2}{2\,c^2}, \qquad (19)$$

Based on the energy conservation law, the radiation energy of elementary particles $E$ with the Compton wavelength $\lambda$

$$E = n\,h\nu = n\,h\,\frac{c}{\lambda}$$

cannot exceed the change in the energy of the body due to the relativistic effect:

$$n\,h\,\frac{c}{\lambda} \leq \Delta m c^2\,.$$

Taking into account equations (14) and (19), we obtain

$$\frac{c}{\lambda} \leq \frac{s\,\Delta V}{4\,\pi\,\rho}. \qquad (20)$$

The maximum value $\Delta V = c$ and inequality (20) takes the form

$$\lambda \geq \frac{4\,\pi\,\rho}{s}. \qquad (21)$$

Constraint (21) can only be satisfied by very low-energy particles. For example, the Compton wavelength of a hypothetical particle graviton $\lambda_g > 1 \cdot 10^{16}$ m [18].

#### 4.4. Russian Experiment on Body Motion in Vacuum

Low-energy particles such as the graviton have energies far beyond the measurement error of the Large Hadron Collider. However, according to the given hypothesis, there is radiation from a stream of low-energy particles with spin. The difference in the number of particles $n_d$ between the number of emitted and absorbed body particles is determined by expressions (15), (17) or (18). In these expressions, there is a change in the body's velocity $\Delta V$ with respect to the detector or a change in height $\Delta H$ and an acceleration $g$, i.e. a relative accelerated motion is required between the detector and the body to register the radiation.

To detect the flow of these particles, the authors carried out an experiment on the basis of the “Scientific Testing Center of the Rocket and Space Industry” (STC RSI) of the State Corporation “Roscosmos” in the vacuum chamber with a volume of 4 m$^3$ (diameter 1.6 m, length 2 m). In the upper part of the vacuum chamber, a mechanical, electrically driven device for the simultaneous release of a cast-iron ball weighing 7.26 kg with a diameter of 11 cm, a bundle of ostrich feathers and a GoPro video camera providing slow-motion shooting with a frequency of 240 frames per second in HD. A garland of ostrich and decorative feathers on a cotton thread (radiation detector) was placed on an independent suspension in the form of a steel wire (removes a static charge) next to the ball's fall path. The vacuum chamber also contained a stationary video camera, a tripod with vertically positioned halogen car lamps, rubber mats to dampen the impact of the ball, and green polyurethane mats to ensure the quality of shooting (Fig. 14)

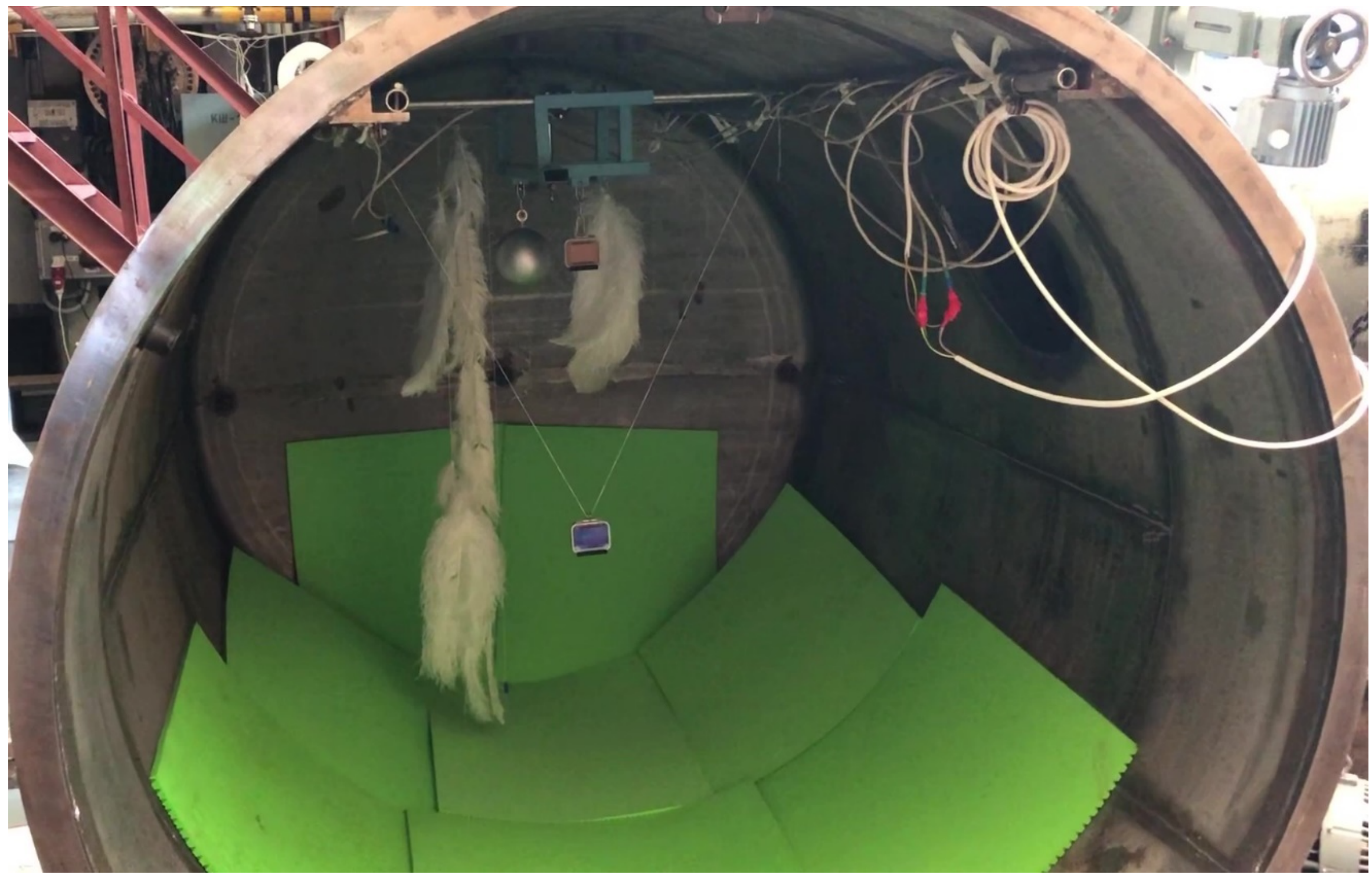

Fig. 14. Equipment for the experiment in the vacuum chamber of the STC RSI

A video recording of the ball falling from a height of 1.1 m made by stationary and mobile GoPro video cameras in vacuum conditions. The backlight consisted of four automotive lamps arranged vertically along a stand. The pressure in the chamber was 0.08 mm Hg. A bunch of ostrich feathers were dropped. At the moment of transition from a suspended state to a state of weightlessness (falling), the fluffs move relative to the ball (there is $\Delta V$). The advantage of this type of motion is that there is no gravity load on the fluff and that the maximum flow of low-energy particles is ensured (21). A clear anomaly in the movement of feathers against the center of the ball is observed in the American experiment described above (Fig. 9). In the presence of radiation towards the ball, the oscillation period of fluffs close to the horizontal plane of the section passing through the mass center of the ball should decrease under the action of a variable radiation flux (15) with a change in $\Delta V$ due to elastic fluff oscillations. In the presence of attraction from the side of the ball, the oscillation period of the fluffs close to the horizontal plane of the section passing through the mass center of the ball should decrease. The presented frames from the GoPro mobile camera show the oscillation of a fluff close to the above plane (Fig. 15). The frequency of its oscillations is higher than that of the others: the fluff manages to make two complete oscillations, while the rest is no more than 1.5. In the first frame (Fig. 15 a), the fluff is deflected by the maximum amplitude from the vertical. In the second frame (Fig. 15 b), the fluff is pressed back to the feather.

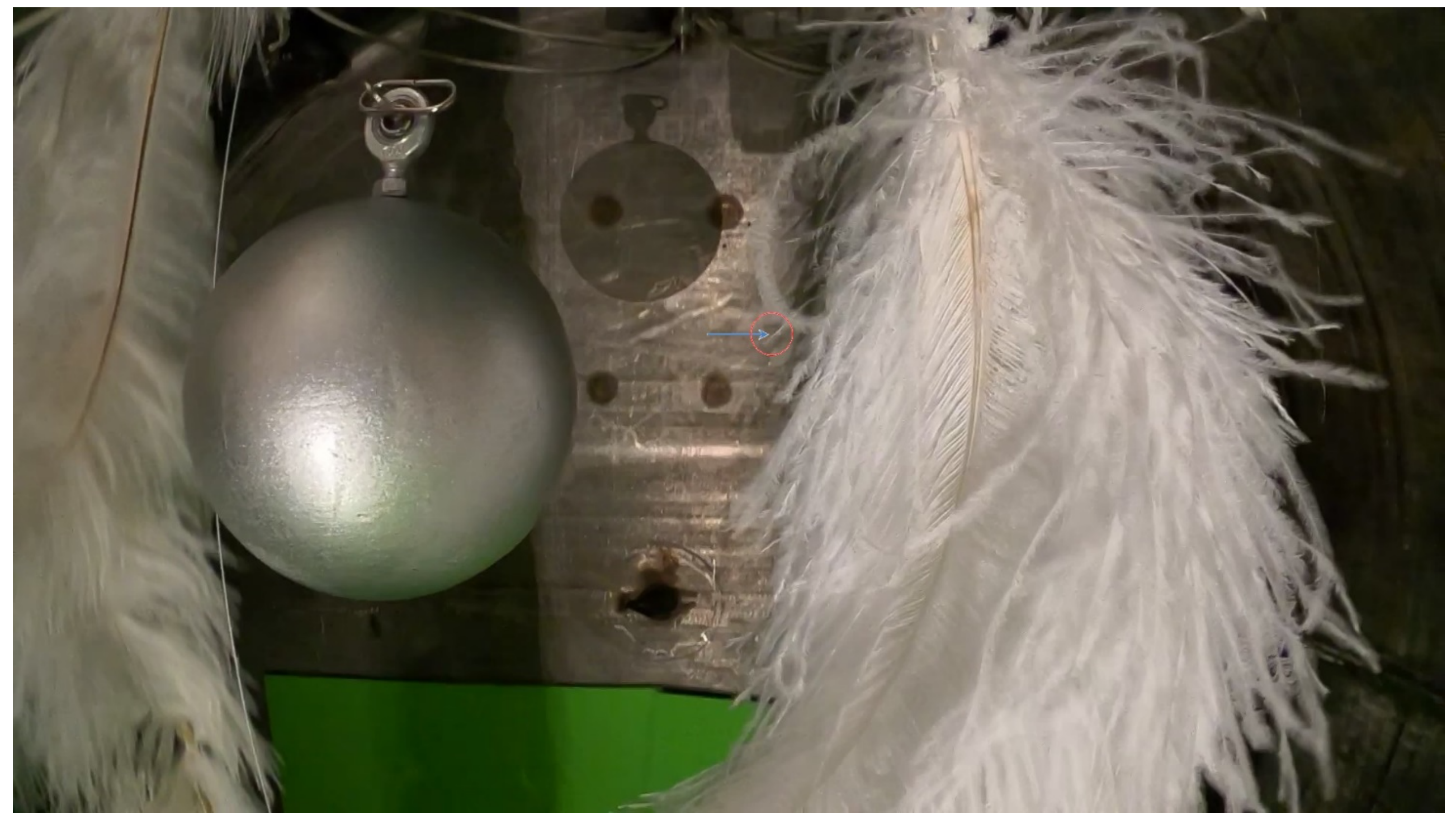

a)

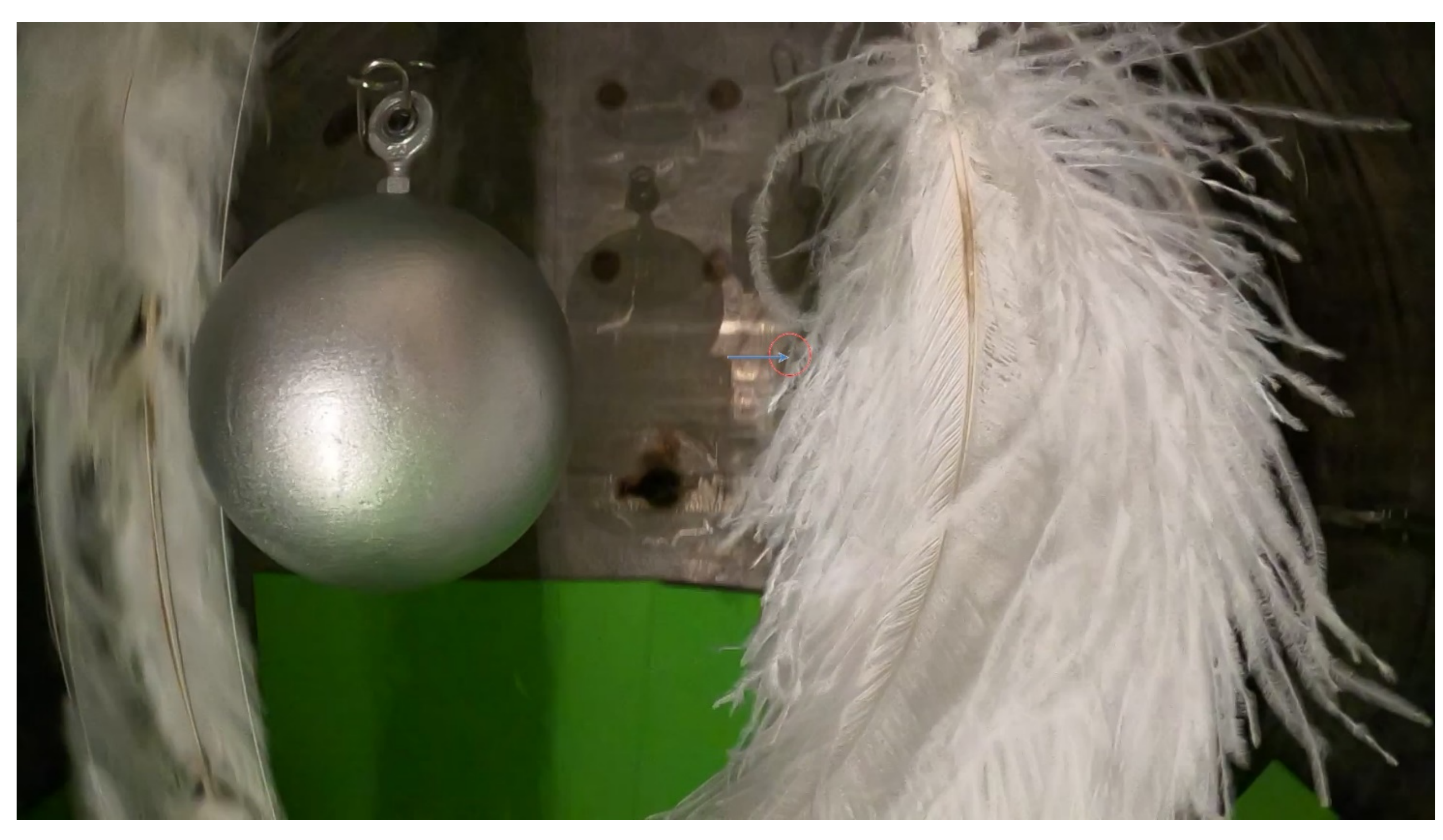

b)

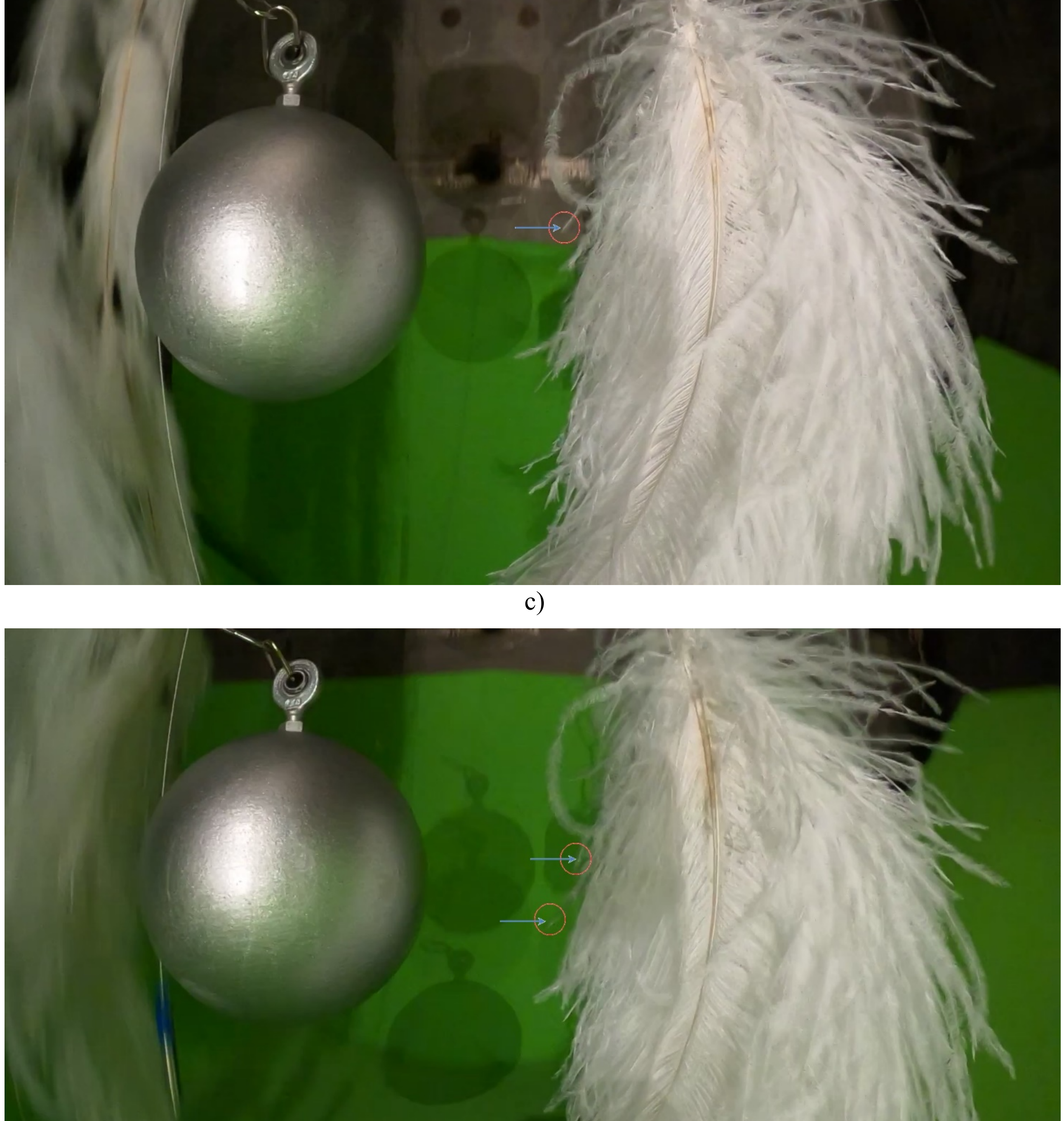

c)

d)

Fig. 15. Video footage of a fall of the ostrich feathers bunch from a falling video camera

In the third frame (Fig. 15c), the fluff again deflects to its maximum amplitude (somewhat less than in the first frame – damped oscillations) towards the ball. In the fourth frame (Fig. 15d), the fluff again tends to the vertical. In the same frame, a new fluff appears for the first time, clearly opposite the center of mass of the ball, slightly lower than the previously considered fluff. Its appearance may be due to the presence of radiation of low-energy particles in the direction of the ball, in particular, this radiation could serve as a trigger when removing it from the engaged position.

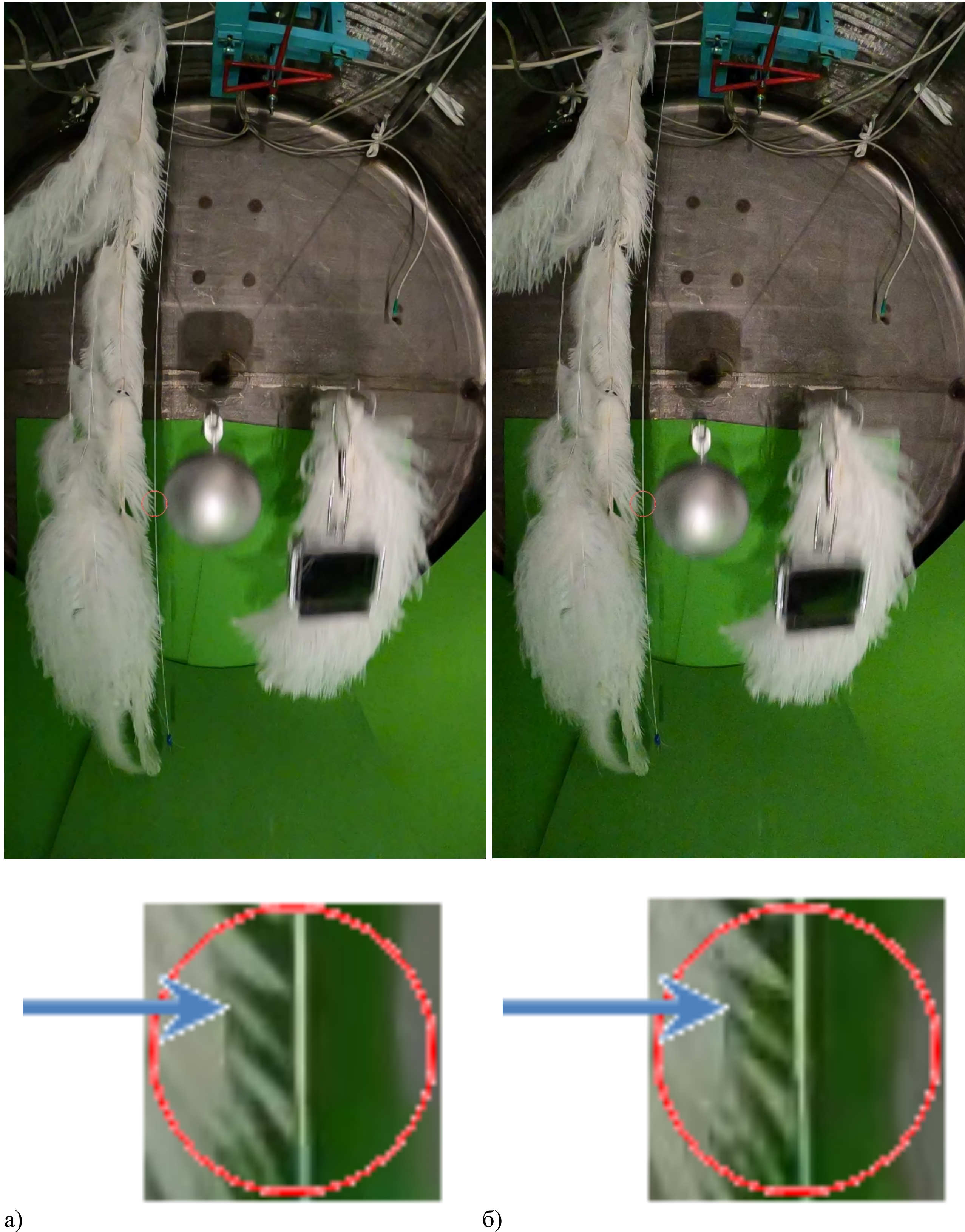

Fig. 16. Frames and fragments of video recordings of the change in the position of the fluff relative to the vertical: a) before the flight of the ball, b) after the flight of the ball

Video footage of the ball falling and the deflection of the garland fluffs from a stationary (suspended on an independent thread) GoPro video camera was obtained (Fig. 16 – 18).

When comparing fragments of frames (Fig. 16) before and after the flight of the ball next to the fluff (indicated by the arrow), its deviation towards the ball is observed. Similar deflections of fluffs are observed at other moments of the ball's falling (Fig. 17 and Fig. 18).

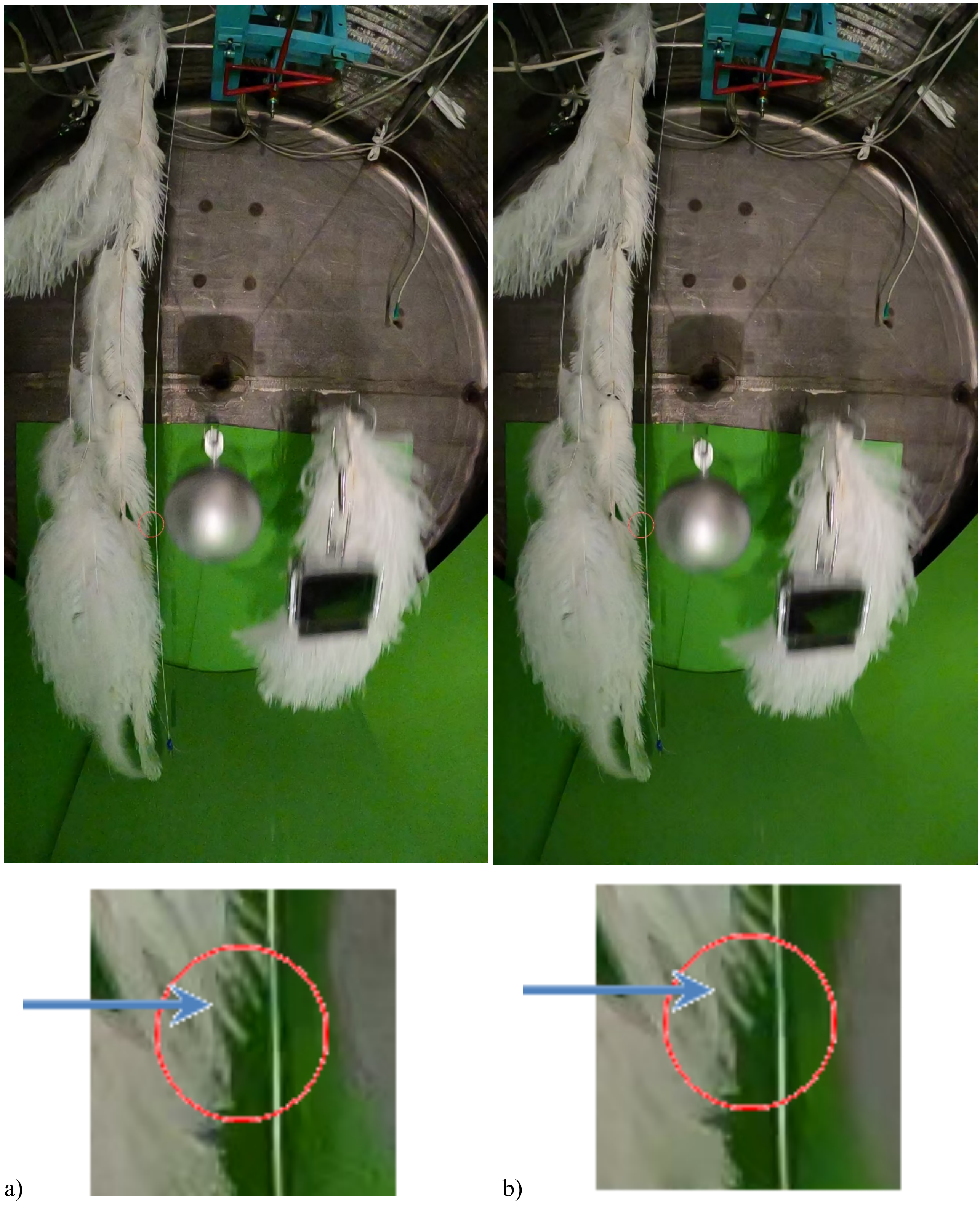

Fig. 18. Frames and fragments of video recordings of the change in the position of the fluff relative to the vertical: a) before the flight of the ball, b) after the flight of the ball

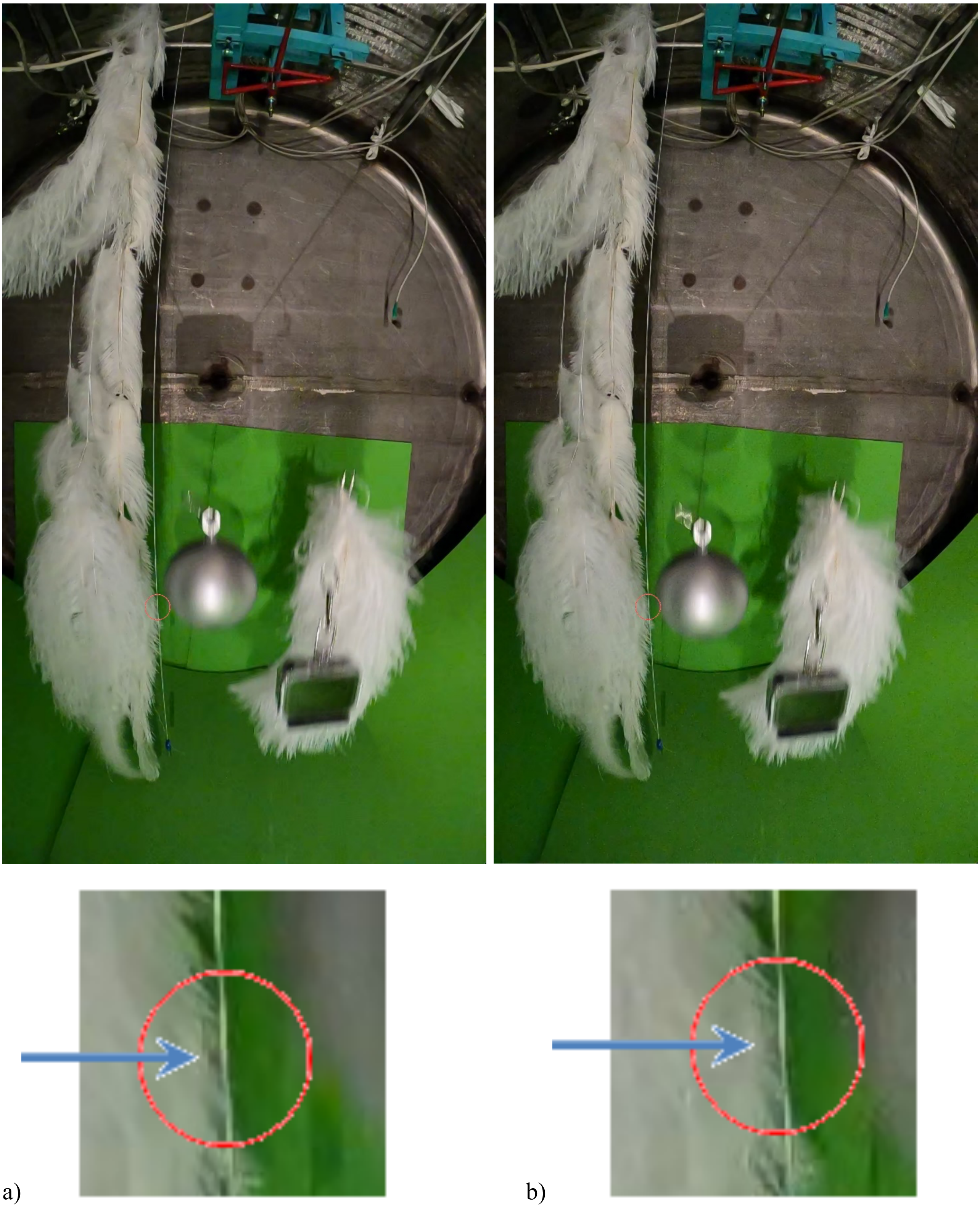

Fig. 18. Frames and fragments of video recordings of changes in the position of fluffs:
a) before the flight of the ball, b) after the flight of the ball

The ball bounced about 30 cm after a strong impact on the damper at the bottom of the vacuum chamber. This circumstance influenced the conditions of the experiment and did not allow testing the hypothesis in terms of body radiation.

The authors plan to carry out more rigorous experiments in a vacuum chamber. From equations (17) and (18) it follows that the maximum effect from the radiation of low-energy particles should be observed at $V_0 = 0$ and high accelerations. Therefore, special attention will be paid to the beginning of the fall of the body ($V_0 = 0$), the moment of rebound upon impact with the damper (large acceleration) and the moment of the body hanging in the upper part of the trajectory after the rebound (change of the direction of radiation of low-energy particles). It is supposed to study the motion of bodies of various shapes.

Numerous examples of fluff movement in a vacuum presented in this section, obtained as a result of each of the four drops of the ball in the framework of an experiment based on the SRC RSP, are consistent with the hypothesis put forward. The authors suggest interested researchers to conduct a similar experiment.

It is important to emphasize that the presented experiments are pilot studies. The observed effects may have alternative explanations, such as the influence of residual gas environment, electrostatic forces, or vibrations. The aim of this work was not to provide irrefutable evidence, but to test the fundamental possibility of detecting anomalous effects.

## 5. On the Technology of Non-Reactive Propulsion Systems

Since the generation of maximum radiation flux from low-energy particles occurs at minimal speeds and high accelerations of a material body, it follows that a device creating thrust based on changes in angular momentum must ensure high-frequency oscillations of the working medium and the reception of useful flows of low-energy elementary particles with spin. The most notable attempts to implement such devices include the EMDrive Thruster [23, 24] and the Mach effect thruster [25].

To develop effective devices, additional experiments are necessary to study the generation of polarized flows of low-energy elementary particles and the possibilities for their reception. Different directions and intensities of low-energy particle fluxes are expected for working bodies of different shapes. Receiving useful low-energy fluxes, forming thrust, is achieved by moving the receiver in the direction of radiation of particles at the time of their generation and subsequent removal of the receiver when changing the direction of radiation.

To achieve motion without overload, it is also important to consider the possibility of creating localized regions in space that are uniformly filled with polarized low-energy elementary particles.

## 6. Advanced Interpretation of the Law of Conservation of Angular Momentum

Taking into account the hypothesis, the movement of a dumbbell with a flywheel (Fig. 3) can be considered as follows:

$$\boldsymbol{K}_{i1} = \boldsymbol{K}_{i0} - \boldsymbol{K}_{r},$$

which does not contradict the law of angular momentum conservation of the system $\boldsymbol{K}$

$$\boldsymbol{K} = \boldsymbol{K}_{i0} + \boldsymbol{K}_{e0} = \boldsymbol{K}_{i1} + \boldsymbol{K}_{e1} = (\boldsymbol{K}_{i0} - \boldsymbol{K}_{r}) + \boldsymbol{K}_{e1} = \boldsymbol{K}_{i1} + (\boldsymbol{K}_{e0} + \boldsymbol{K}_{r}),$$

where $\boldsymbol{K}_{i0}$, $\boldsymbol{K}_{i1}$ – the initial (at time $t_0$) and final (at time $t_1$) vectors of the internal angular momentum relative to the mass center of the dumbbell *C*,

$\boldsymbol{K}_{e0}$, $\boldsymbol{K}_{e1}$ – the initial (at time $t_0$) and final (at time $t_1$) angular momentum vectors of the mass center of the dumbbell *C*,

$\boldsymbol{K}_r$ – the vector of the angular momentum of radiation (or absorption) of elementary particles over time $\Delta t = t_1 - t_0$ due to the accelerated rotation of the flywheel.

This representation of the angular momentum of a body allows a more in-depth explanation of the phenomenon of radial displacement of the mass center *C* in the central gravitational field.

For direct verification of the proposed mechanism, it is advisable to conduct a specialized space experiment according to the scheme presented in Fig. 4. Such research would not only detect the predicted directional emission ("jet"), but also analyze its influence on vacuum fluctuation density and electron-positron pair production processes. This, in turn, would open a path to understanding the fundamental mechanism of jet formation from accretion disks in astrophysics.

Indirect confirmation of the hypothesis is found in observed phenomena of stellar dynamics. The pattern of polar jet flows, where matter is ejected along the rotation axis of a cosmic object's accretion disk due to internal dynamic processes, demonstrates striking similarity to the emission arising from changes in kinetic momentum (Fig. 6).

The proposed approach offers a new interpretation of one of the key problems in modern astrophysics - the discrepancy between observed rotational velocities of matter in spiral galaxy disks and predictions of classical dynamics that account only for visible mass. Within the hypothesis, changes in the kinetic momentum of matter near the galactic center are accompanied by emission of low-energy particles in the direction perpendicular to the galactic plane. Under the influence of the galaxy's gravitational field, this scattered radiation forms a structure analogous to magnetic field lines. Thus, the galaxy becomes permeated by polarized flows of low-energy particles, which may exert additional influence on stellar dynamics and explain the observed rotation curve profiles without invoking the concept of dark matter.

This explanation preserves the formal characteristics of the phenomenon - the constancy of orbital velocities of stars across a wide range of distances from the galactic center - while proposing a fundamentally different physical mechanism for it, based on polarization effects in the spin structure of spacetime.

## 7. Towards a Theory of Gravity

The necessity of reconciling the quantum description of matter with the geometric description of spacetime remains one of the fundamental problems in modern physics. Despite significant efforts, a complete and consistent theory of quantum gravity has not yet been developed. Existing approaches, such as string theory [26] and loop quantum gravity [27], face challenges in experimental verification, particularly in the domain of low-energy gravitational effects [28, 29].

In contrast to these directions, this work proposes a novel approach to explaining the phenomenon of gravity at Planck scales, consistent with General Relativity, based on a mechanistic interpretation of gravity and inertia through the hypothesis of spacetime spin polarization. According to this hypothesis, spacetime constitutes a dynamic medium composed of low-energy carrier particles (conventionally termed "gravitons"), existing in two states: free and bound. Gravitational attraction is explained by the density gradient of free particles. A key distinction from standard quantum field theory is that gravitational attraction is explained not by the exchange of virtual particles, but by the difference in the density of carrier particles in the free state. In the vicinity of massive bodies, the density of free particles decreases, creating a pressure gradient that induces the attraction effect.

This approach allows for a mechanistic interpretation of spacetime geometry: curvature and geodesic lines are formed by the distribution and state of the carrier particles, analogous to how tightening threads into knots forms the structure of a fabric. Gravitational waves in this model correspond to density waves within the particle medium. The proposed approach also offers a novel explanation for the dark matter phenomenon [30]: observed anomalies in galactic rotation curves may result from the global spin polarization of low-energy particle fluxes generated by galactic rotation.

The new perspective on the nature of inertia presents an original alternative to both the Machian interpretation and the standard General Relativity approach. Inertia arises as a consequence of asymmetry in the absorption/emission of spin-polarized particles during accelerated motion.

A key advantage of the hypothesis is its experimental testability at accessible energy levels. The directional particle fluxes predicted by the model can be detected under laboratory conditions, as supported by the pilot experiments involving falling bodies in a vacuum described in Section 4. Significant prospects in this direction are demonstrated by studies that have detected particles with spin 2 [31], which can be viewed as indirect confirmation of the proposed model.

These results indicate the fundamental possibility of detecting spacetime spin polarization in the vicinity of accelerating bodies, opening a new direction in experimental gravity. Unlike traditional methods [32], such as precision measurements of gravitational interaction [33-36], detection of gravitational waves [37, 38], and indirect tests of quantum gravity [39], the proposed approach focuses on investigating polarization effects.

The primary challenge in experimental implementation lies in the high scattering probability of low-energy particles due to their large Compton wavelength. However, observed anomalous effects in astrophysical dynamics suggest the existence of a sufficient spin relaxation time for detecting polarized fluxes of "graviton"-type particles using modern scientific equipment.

A promising method for generating directional fluxes involves creating polarization in the form of a "beam" or "jet" (Fig. 6) by maintaining a "dumbbell-flywheel" system at an angle to the local vertical within a central gravitational field. The authors recommend that traditional gravitational experiments account for potential anomalies caused by body motion, which may be interpreted as manifestations of low-energy particle flux polarization.

If experimentally confirmed, the hypothesis regarding the emission/absorption of low-energy particles by material bodies will establish a connection between gravity and the physics of the microcosm within the framework of a mechanistic model of quantum particle interaction, as well as offer a new explanation for the phenomenon of wave-particle duality.

The proposed hypothesis is consistent, at Planck scales, with the fundamental laws of physics: the law of conservation of momentum, the law of conservation of angular momentum, the law of conservation of energy, and the law of conservation of the center of mass position.

## 8. Conclusion

Theoretically, the possibility and energy feasibility of implementing the idea of creating thrust based on changes in angular momentum have been proven for the development of transport objects on new physical principles. The proposed hypothesis, along with the examples and experiments provided, gives grounds for the formation of new physical concepts regarding the mechanistic nature of elementary particle interactions. The results obtained can be used in experiments aimed at searching for low-energy elementary particles. The practical implementation of this idea requires further fundamental research in the following areas:

1. Experimental Verification. A more rigorous experiment is needed to register streams of low-energy particles with spin. Methods must be developed to distinguish between the emission and absorption of particles depending on the state of motion of the body.
2. Analysis of Existing Experiment Results. Reevaluating data obtained from previously conducted experiments on gravity and gravitational waves in light of the possible influence of polarized low-energy particles. This may help uncover new aspects of the interaction between matter and gravity.
3. Modeling Interactions. Creating theoretical models that describe the transition of low-energy particles from a free state to a bound state and back. This may include numerical simulations that take into account various states of particles and their influence on the geometry of spacetime, as well as on the motion of elementary particles in the form of a wave packet.
4. Interdisciplinary Research. Collaborating with physicists, mathematicians, and engineers to create a comprehensive approach to investigating this topic. This could involve the use of new technologies and methods from various fields of science.
5. Development of New Theories. Based on the data obtained and experimental results, it may be necessary to formulate new theories that integrate quantum mechanics and gravity, considering the influence of the polarization of low-energy particles on gravitational interactions.

These research directions could significantly deepen our understanding of gravity and promise both scientific breakthroughs and practical improvements in humanity's capabilities for space exploration.

**Acknowledgments**

This paper has been supported by the Ministry of Science and Higher Education of the Russian Federation under Agreement No. FSSF-2024-0005.

**Author Declarations**

The authors have no conflicts to disclose.

**Data availability**

The data that support the findings of this study are available from the corresponding author upon reasonable request.

**Appendix. Equations of a rigid dumbbell motion in the Earth central gravitational field**

Let us define expressions (4) and (6).

Modules of gravity forces $\boldsymbol{G_1}$ and $\boldsymbol{G_2}$

$$G_1 = \mu_0 \frac{m_1}{r_1^2}; \quad G_2 = \mu_0 \frac{m_2}{r_2^2}. \tag{22}$$

Projections of gravity forces $\boldsymbol{G_1}$ and $\boldsymbol{G_2}$ on the *Cx* axis (Fig. 2):

$$G_{1x} = -G_1 \sin\alpha \ ; \quad G_{2x} = G_2 \sin\beta . \tag{23}$$

By the sine theorem

$$\frac{\sin\alpha}{D_1} = \frac{\sin(\pi/2+\varepsilon)}{r_1}; \quad \frac{\sin\beta}{D_2} = \frac{\sin(\pi/2-\varepsilon)}{r_1};$$

$$\sin\alpha = \frac{D_1}{r_1}\cos\varepsilon; \quad \sin\beta = \frac{D_2}{r_2}\cos\varepsilon; \tag{24}$$

Projections of gravity forces $\boldsymbol{G_1}$ and $\boldsymbol{G_2}$ on the *Cy* axis (Fig. 2):

$$G_{1y} = -G_1 \cos\alpha; \quad G_{2y} = -G_2 \cos\beta; \tag{25}$$

where

$$\cos\alpha = \frac{r + D_1 \sin\varepsilon}{r_1}; \quad \cos\beta = \frac{r - D_2 \sin\varepsilon}{r_2}. \tag{26}$$

From (3), (22), (23) and (24) and taking into account $m_1 D_1 = m_2 D_2$ (7) we successively obtain the projection of the main force vector onto the axis *Cx*:

$$F_{Cx} = -G_1 \sin\alpha + G_2 \sin\beta;$$

$$F_{Cx} = -\mu_0 \frac{m_1}{r_1^2}\frac{D_1}{r_1}\cos\varepsilon + \mu_0 \frac{m_2}{r_2^2}\frac{D_2}{r_2}\cos\varepsilon;$$

$$F_{Cx} = \mu_0 m_1 D_1 \cos\varepsilon \left(\frac{1}{r_2^3} - \frac{1}{r_1^3}\right). \tag{27}$$

From equations (5), (22), (25) and (26) we obtain the projection of the main force vector onto the axis *Cy*:

$$F_{Cy} = -G_1 \cos\alpha - G_2 \cos\beta;$$

$$F_{Cy} = -\mu_0 \frac{m_1}{r_1^2} \frac{(r + D_1 \sin\varepsilon)}{r_1} - \mu_0 \frac{m_2}{r_2^2} \frac{(r - D_2 \sin\varepsilon)}{r_2};$$

$$F_{Cy} = -\mu_0 \frac{m_1 r}{r_1^3} - \mu_0 \frac{m_1 D_1 \sin\varepsilon}{r_1^3} - \mu_0 \frac{m_2 r}{r_2^3} + \mu_0 \frac{m_2 D_2 \sin\varepsilon}{r_2^3};$$

$$F_{Cy} = -\mu_0 r \left( \frac{m_1}{r_1^3} + \frac{m_2}{r_2^3} \right) + \mu_0 m_1 D_1 \sin\varepsilon \left( \frac{1}{r_2^3} - \frac{1}{r_1^3} \right);$$

$$F_{Cy} = -\mu_0 r \left( \frac{m_1}{r_1^3} + \frac{m_1}{r^3} - \frac{m_1}{r^3} + \frac{m_2}{r_2^3} + \frac{m_2}{r^3} - \frac{m_2}{r^3} \right) + \mu_0 m_1 D_1 \sin\varepsilon \left( \frac{1}{r_2^3} - \frac{1}{r_1^3} \right);$$

$$F_{Cy} = -\mu_0 \frac{m}{r^2} + \mu_0 r m_1 \left( \frac{1}{r^3} - \frac{1}{r_1^3} \right) - \mu_0 r m_2 \left( \frac{1}{r_2^3} - \frac{1}{r^3} \right) + \mu_0 m_1 D_1 \sin\varepsilon \left( \frac{1}{r_2^3} - \frac{1}{r_1^3} \right). \qquad (28)$$

By the cosine theorem

$$\frac{1}{r_2^3} - \frac{1}{r_1^3} = \frac{1}{r^3} \left( \frac{r^2 + D_2^2 - 2rD_2 \cos(\pi/2 - \varepsilon)}{r^2} \right)^{-3/2}$$
$$- \frac{1}{r^3} \left( \frac{r^2 + D_1^2 - 2rD_1 \cos(\pi/2 + \varepsilon)}{r^2} \right)^{-3/2} =$$
$$= \frac{1}{r^3} \left[ \left( 1 - \frac{2D_2 \sin\varepsilon}{r} + \frac{D_2^2}{r^2} \right)^{-3/2} - \left( 1 + \frac{2D_1 \sin\varepsilon}{r} + \frac{D_1^2}{r^2} \right)^{-3/2} \right]$$

and applying the Newton binomial formula with preservation of the second-order terms in the expansion, assuming ($D \ll r$)

$$\left( 1 - \frac{2D_2 \sin\varepsilon}{r} + \frac{D_2^2}{r^2} \right)^{-3/2} \cong 1 + \frac{3D_2 \sin\varepsilon}{r} - \frac{3D_2^2}{2r^2};$$

$$\left( 1 + \frac{2D_1 \sin\varepsilon}{r} + \frac{D_1^2}{r^2} \right)^{-3/2} \cong 1 - \frac{3D_1 \sin\varepsilon}{r} - \frac{3D_1^2}{2r^2};$$

obtain

$$\frac{1}{r_2^3} - \frac{1}{r_1^3} \cong \frac{3D \sin\varepsilon}{r^4} + \frac{3(D_1^2 - D_2^2)}{2r^5}. \qquad (29)$$

Analogically

$$\frac{1}{r^3} - \frac{1}{r_1^3} = \frac{1}{r^3} \left[ 1 - \left( 1 + \frac{2D_1 \sin\varepsilon}{r} + \frac{D_1^2}{r^2} \right)^{-3/2} \right];$$

$$\frac{1}{r^3} - \frac{1}{r_1^3} \cong \frac{3D_1 \sin\varepsilon}{r^4} + \frac{3D_1^2}{2r^5}. \qquad (30)$$

As well as

$$\frac{1}{r_2^3}-\frac{1}{r^3}=\frac{1}{r^3}\left[\left(1-\frac{2D_2\sin\varepsilon}{r}+\frac{D_2^2}{r^2}\right)^{-3/2}-1\right];$$

$$\frac{1}{r_2^3}-\frac{1}{r^3}\cong\frac{3D_2\sin\varepsilon}{r^4}-\frac{3D_2^2}{2r^5}. \tag{31}$$

Taking into account (29), equation (27) will take the form

$$F_{Cx}=\mu_0 m_1 D_1\cos\varepsilon\frac{3D\sin\varepsilon}{r^4}+\mu_0 m_1 D_1\cos\varepsilon\frac{3(D_1^2-D_2^2)}{2r^5}.$$

Without taking into account the term of the highest order of smallness, we obtain the equation (4)

$$F_{Cx}=\mu_0 m_1 D_1\cos\varepsilon\frac{3D\sin\varepsilon}{r^4};$$

$$F_{Cx}=\frac{3}{2}\mu_0\frac{mD^2}{r^4}\frac{\eta}{(1+\eta)^2}\sin 2\varepsilon\,.$$

Taking into account (29), (30), (31) equation (28) will take the form

$$F_{Cy}=-\mu_0\frac{m}{r^2}+\mu_0 r m_1\frac{3D_1\sin\varepsilon}{r^4}-\mu_0 r m_2\frac{3D_2\sin\varepsilon}{r^4}+\mu_0\frac{3(m_1D_1^2+m_2D_2^2)}{2r^4}$$
$$+\mu_0 m_1 D_1\sin\varepsilon\frac{3D\sin\varepsilon}{r^4}+\mu_0 m_1 D_1\sin\varepsilon\frac{3(D_1^2-D_2^2)}{2r^5}.$$

Without taking into account the term of the highest order of smallness, we obtain the equation (6)

$$F_{Cy}=-\mu_0\frac{m}{r^2}+\mu_0\frac{3(m_1D_1^2+m_2D_2^2)}{2r^4}+\mu_0 m_1 D_1\sin\varepsilon\frac{3D\sin\varepsilon}{r^4};$$

$$F_{Cy}=-\mu_0\frac{m}{r^2}+\frac{3}{2}\mu_0\frac{mD^2}{r^4}\frac{\eta}{(1+\eta)^2}+3\mu_0\frac{mD^2}{r^4}\frac{\eta}{(1+\eta)^2}\sin^2\varepsilon;$$

$$F_{Cy}=-\mu_0\frac{m}{r^2}+3\mu_0\frac{mD^2}{r^4}\frac{\eta}{(1+\eta)^2}\left(\frac{1}{2}+\sin^2\varepsilon\right).$$

Let us define expression (8) for the moment $M_C(\boldsymbol{G_1},\boldsymbol{G_2})$. According to Varignon's theorem, taking into account (2), (7), (22), (24), (26) and (29), we sequentially obtain the equation for $M_C$:

$$M_C=G_1\sin\alpha\,D_1\sin\varepsilon-G_1\cos\alpha\,D_1\cos\varepsilon+G_2\sin\beta\,D_2\sin\varepsilon+G_2\cos\beta\,D_2\cos\varepsilon\,;$$

$$M_C=G_1D_1(\sin\alpha\sin\varepsilon-\cos\alpha\cos\varepsilon)+G_2D_2(\sin\beta\sin\varepsilon+\cos\beta\cos\varepsilon);$$

$$M_C=\mu_0\frac{m_1}{r_1^2}D_1(\sin\alpha\sin\varepsilon-\cos\alpha\cos\varepsilon)+\mu_0\frac{m_2}{r_2^2}D_2(\sin\beta\sin\varepsilon+\cos\beta\cos\varepsilon);$$

$$M_C=\mu_0 m_1 D_1\left[\frac{\sin\alpha\sin\varepsilon-\cos\alpha\cos\varepsilon}{r_1^2}+\frac{\sin\beta\sin\varepsilon+\cos\beta\cos\varepsilon}{r_2^2}\right];$$

$$M_C=\mu_0 m_1 D_1\left(\frac{D_1}{r_1^3}\cos\varepsilon\sin\varepsilon-\frac{r+D_1\sin\varepsilon}{r_1^3}\cos\varepsilon+\frac{D_2}{r_2^3}\cos\varepsilon\sin\varepsilon\right.$$
$$\left.-\frac{r-D_2\sin\varepsilon}{r_2^3}\cos\varepsilon\right);$$

$$M_C=\mu_0 m_1 D_1 r\cos\varepsilon\left(\frac{1}{r_2^3}-\frac{1}{r_1^3}\right);$$

$$M_C \cong \mu_0 m_1 D_1 \cos\varepsilon \frac{3D \sin\varepsilon}{r^3};$$

$$M_C = \frac{3}{2}\mu_0 \frac{m_1 D_1 D}{r^3}\sin 2\varepsilon;$$

$$M_C = \frac{3}{2}\mu_0 \frac{m_1 D_1 (D_1 + D_2)}{r^3}\sin 2\varepsilon;$$

$$M_C = \frac{3}{2}\mu_0 \frac{(m_1 D_1^2 + m_2 D_2^2)}{r^3}\sin 2\varepsilon;$$

$$M_C = \frac{3}{2}\mu_0 \frac{J_D}{r^3}\sin 2\varepsilon;$$

where

$$J_D = m_1 D_1^2 + m_2 D_2^2 .$$

Let us define the equations of motion of the dumbbell's mass center.

The fundamental law of dynamics for the motion of the dumbbell's mass center *C* in an absolute coordinate system

$$m\boldsymbol{a} = \boldsymbol{F}_C ; \tag{32}$$

$\boldsymbol{a}$ – the acceleration of the dumbbell's mass center.

(32) in the polar coordinate system $(r, \vartheta)$ (Fig. 2):

$$m(\ddot{r} - \dot{\vartheta}^2 r) = F_{Cy} ;$$

$$m(r\ddot{\vartheta} + 2\dot{r}\dot{\vartheta}) = -F_{Cx} .$$

Taking into account (4) and (6), we obtain the equations of motion of the dumbbell's mass center *C*

$$\ddot{r} - \dot{\vartheta}^2 r = -\frac{\mu_0}{r^2} + 3\mu_0 \frac{D^2}{r^4}\frac{\eta}{(1+\eta)^2}\left(\frac{1}{2} + \sin^2\varepsilon\right); \tag{33}$$

$$r\ddot{\vartheta} + 2\dot{r}\dot{\vartheta} = -\frac{3}{2}\mu_0 \frac{D^2}{r^4}\frac{\eta}{(1+\eta)^2}\sin 2\varepsilon. \tag{34}$$

Let us determine the equation of the dumbbell's motion relative to the mass center. For $J_D = const$

$$M_C = J_D(\ddot{\varepsilon} + \ddot{\vartheta}).$$

Then, taking into account (8), the equation of angular motion of the dumbbell

$$\ddot{\varepsilon} = \frac{3}{2}\mu_0 \frac{\sin 2\varepsilon}{r^3} - \ddot{\vartheta}. \tag{35}$$

The system of equations (33), (34), (35) with a circular flywheel of mass $m_J$ supporting the angle ε takes the form (9):

$$\ddot{r} - \dot{\vartheta}^2 r = -\frac{\mu_0}{r^2} + 3\mu_0 \frac{D^2}{r^4}\frac{\eta}{(1+\eta)^2}\left(\frac{1}{2} + \sin^2\varepsilon\right)\frac{m}{(m + m_J)};$$

$$r\ddot{\vartheta} + 2\dot{r}\dot{\vartheta} = -\frac{3}{2}\mu_0 \frac{D^2}{r^4} \frac{\eta}{(1+\eta)^2} \frac{m}{(m+m_J)} \sin 2\varepsilon;$$

$$\varepsilon = const\,.$$